\documentclass[lettersize,journal]{IEEEtran}
\usepackage{amsmath,amssymb,amsfonts}
\usepackage[linesnumbered,ruled]{algorithm2e}
\usepackage{array}
\usepackage{textcomp}
\usepackage{stfloats}
\usepackage{url}
\usepackage{color}
\usepackage{verbatim}
\usepackage{graphicx}
\usepackage{tabularx}
\usepackage{multirow}
\usepackage{array}
\usepackage{booktabs}
\usepackage{graphicx}
\usepackage{parskip}

\usepackage{amsthm}
\newtheorem{Definition}{Definition}
\usepackage[numbers]{natbib}
\usepackage{bm}
\usepackage{mathtools}
\DeclareMathOperator*{\argmax}{arg\,max}
\usepackage{subfigure}
\usepackage{makecell}
\usepackage[table]{xcolor}
\usepackage{multirow} 
\usepackage{enumitem}
\usepackage{bibspacing}
\graphicspath{ {./fig/} }
\usepackage{flushend}

\def\BibTeX{{\rm B\kern-.05em{\sc i\kern-.025em b}\kern-.08em
    T\kern-.1667em\lower.7ex\hbox{E}\kern-.125emX}}
\usepackage{balance}
\begin{document}
\title{Task-Specific Trust Evaluation for Multi-Hop Collaborator Selection via GNN-Aided \\Distributed Agentic AI}

 \author{
 Botao~Zhu,~\IEEEmembership{Member,~IEEE}, Xianbin~Wang,~\IEEEmembership{Fellow,~IEEE}, and Dusit~Niyato,~\IEEEmembership{Fellow,~IEEE
 }
 \thanks{
 B. Zhu and X. Wang are with the Department of Electrical and Computer Engineering, Western University, London, Canada N6A 5B9 (Emails: \{bzhu88, xianbin.wang\}@uwo.ca).
 
 D. Niyato is with the School of Computer Science and Engineering, Nanyang Technological University, Singapore (Email: dniyato@ntu.edu.sg).}}


\maketitle

\begin{abstract}   
    The success of collaborative task completion among networked devices hinges on the effective selection of trustworthy collaborators. However, accurate task-specific trust evaluation of multi-hop collaborators can be extremely complex. The reason is that their trust evaluation is determined by a combination of diverse trust-related perspectives with different characteristics, including historical collaboration reliability, volatile and sensitive conditions of available resources for collaboration, as well as continuously evolving network topologies. To address this challenge, this paper presents a graph neural network (GNN)-aided distributed agentic AI (GADAI) framework, in which different aspects of devices’ task-specific trustworthiness are separately evaluated and jointly integrated to facilitate multi-hop collaborator selection. GADAI first utilizes a GNN-assisted model to infer device trust from historical collaboration data. Specifically, it employs GNN to propagate and aggregate trust information among multi-hop neighbours, resulting in more accurate device reliability evaluation. Considering the dynamic and privacy-sensitive nature of device resources, a privacy-preserving resource evaluation mechanism is implemented using agentic AI. Each device hosts a large AI model-driven agent capable of autonomously determining whether its local resources meet the requirements of a given task, ensuring both task-specific and privacy-preserving trust evaluation. By combining the outcomes of these assessments, only the trusted devices can coordinate a task-oriented multi-hop cooperation path through their agents in a distributed manner. Experimental results show that our proposed GADAI outperforms the comparison algorithms in planning multi-hop paths that maximize the value of task completion.
\end{abstract}


\section{Introduction}

\subsection{Motivation}
As applications and interconnected systems become increasingly complex, individual devices are often unable to handle computationally intensive tasks independently due to their limited processing power and energy constraints. To overcome this challenge, researchers have begun to explore distributed resource scheduling, where tasks are offloaded to devices with more computational power in the system for execution via multi-hop relaying~\cite{9252973}. For example, in vehicular networks, autonomous vehicles can offload computational tasks to nearby roadside units~\cite{9978606}, while in the industrial Internet of Things systems, sensor nodes can delegate data processing tasks to edge gateways~\cite{9833361}. To ensure effective task completion, the selection of relay and computing devices with high stability and reliability is crucial. 
Various mechanisms, such as access control mechanisms, reputation-based approaches, and authentication and authorization protocols, have been employed to select collaborators~\cite{9831429}. While these mechanisms enhance security and stability to a certain extent, they are insufficient to ensure the reliability of devices in dynamic systems.

Trust is becoming an essential means of assessing collaborator reliability in collaborative systems. In such systems, trust is typically defined as the task owner's confidence in a collaborator's ability to perform tasks, which is evaluated based on historical behaviours, such as packet loss rate, task completion rate, and session interruption frequency~\cite{Zhu2025RapidContinuousTrust}. However, in increasingly complex and dynamic collaborative systems, device trust is influenced not only by historical behaviours but also by factors such as resource status and network topology~\cite{chain_of_trsut}. Therefore, this paper aims to explore how to integrate multidimensional and diverse trust-related data in complex system environments to achieve a more accurate trust evaluation for collaborators, thereby providing a solid foundation for reliable multi-hop cooperation.

\vspace{-0.1 in}
\subsection{Key Technical Challenges}
Building trusted multi-hop collaboration paths must focus on the core objective of task completion, which involves task-specific trust evaluation of devices and multi-hop path planning oriented toward successful task execution. Nevertheless, device trust evaluation is determined by a combination of diverse trust-related perspectives, including the reliability of historical collaborations, the volatility and sensitivity of available collaborative resources, and the dynamically changing network topology, which brings the following challenges for accurate task-specific device trust evaluation and task-oriented multi-hop path planning.

\textit{\textbf{How to design an effective mechanism to handle diverse trust-related perspectives?}} Accurate task-specific trust evaluation of devices requires collecting their data across multiple dimensions, such as historical collaboration records and available resource information. Each of these dimensions, which represents a trust-related perspective, exhibits distinct characteristics. For example, historical collaboration data is relatively stable and can be leveraged for long-term reliability inference, whereas device resources are highly dynamic and often cannot be shared due to privacy constraints. Existing studies typically overlook the multi-dimensional nature of these data and adopt general trust evaluation frameworks to process all dimensions uniformly~\cite{11096939}. As a result, these approaches struggle to provide accurate trust assessment results. To overcome these limitations, this work proposes a separate evaluation and integrated decision-making framework that separately assesses different aspects of task-specific device trustworthiness and jointly integrates these results to facilitate multi-hop collaborator selection.

\textit{\textbf{How to evaluate historical collaboration reliability?}} Historical collaboration data encompass not only direct collaborations between one-hop neighbours, such as task forwarding and computing, but also indirect collaborations arising through multi-hop neighbours. Some researchers collect multi-dimensional historical data, such as task completion rates, packet loss rates, computation delays, and causes of task failure, and then use predefined rules or machine learning methods to infer device trust~\cite{10422726},~\cite{8769947}. However, these studies rely solely on the direct collaborations of one-hop neighbours for trust assessment~\cite{7194748}, \cite{9335844}, which limits their ability to capture multi-hop dependencies, resulting in inaccurate trust assessments. To fully leverage historical collaboration information and capture complex trust dependencies, it is necessary to develop mechanisms that can represent intricate inter-device collaborations and propagate trust information across the network topology. Recent studies have shown that graph structures are well-suited for representing complex collaborations, and graph neural networks (GNNs) enable multi-hop message propagation~\cite{Wu2021Comprehensive}, making them promising tools for addressing this challenge. Therefore, this research aims to develop a graph model for representing device collaborations and to employ a GNN model to aggregate and propagate trust information across multi-hop neighbours, achieving a more accurate evaluation of historical reliability.

\textbf{\textit{How to evaluate volatile and privacy-preserving resources?}} Device resources, such as CPU, energy, memory, and bandwidth, play a crucial role in determining whether a device can reliably fulfill a given task~\cite{chain_of_trsut}, making resource assessment an essential component of trust evaluation. Existing studies typically employ general-purpose metrics, such as resource capacity, workload, or energy consumption models, to assess resource reliability~\cite{10320384},~\cite{10251781}. However, these studies, which rely on machine learning methods or predefined rules, struggle to reflect the time-varying characteristics of resource trust and lack effective mechanisms for protecting resource privacy. In addition, the trustworthiness of a device's resources may vary considerably across different tasks, yet these methods fail to account for the task-specific nature of resource trust. Agentic AI, characterized by task-specific reasoning, autonomous decision-making, and environmental adaptability, offers a promising solution to these challenges. In this work, we explore enabling devices to perform task-specific and privacy-preserving trust evaluations via agentic AI.

\textbf{\textit{How to plan a task-oriented multi-hop path?}}
The primary objective of constructing multi-hop cooperative paths among devices is to enable efficient task completion. Some studies have applied swarm intelligence algorithms, such as ant colony optimization and particle swarm optimization~\cite{6331686},~\cite{10838540}, to plan multi-hop cooperative paths in wireless systems, while others have leveraged machine learning methods to optimize route selection~\cite{10812999},~\cite{10577995}. However, in distributed systems, inter-device connections are highly dynamic, rendering these centralized or static approaches less effective. Recently, multi-agent systems have emerged as a promising solution, where agents endowed with autonomous decision-making and collaborative capabilities achieve global objectives through local perception and collaboration. Building on this concept, this study proposes a task-oriented multi-hop path planning framework, in which large AI model (LAM)-driven agents empower devices with dynamic decision-making and task-oriented collaboration, facilitating multi-agent cooperative path construction in distributed environments.

\vspace{-0.05 in}
\subsection{Contributions}
To overcome the aforementioned challenges, this paper proposes a GNN-aided distributed agentic AI (GADAI) framework, in which different aspects of devices’ task-specific trustworthiness are separately assessed and then integrated to support multi-hop collaborator selection. Concretely, the GNN-assisted trust evaluation model is designed to infer the historical reliability of devices. It first constructs a historical collaboration graph to capture trust dependencies among devices. The GNN model then propagates and aggregates trust information over the graph, allowing each device to integrate trust information from multi-hop neighbours for a more accurate evaluation of historical reliability. Given the dynamic and privacy-sensitive nature of device resources, we propose a resource evaluation scheme based on agentic AI. Each device is equipped with an LAM-driven agent, which autonomously assesses whether local resources satisfy specific task requirements, thereby achieving task-specific and privacy-preserving resource trust evaluation. By integrating the results from these evaluations, only trusted devices coordinate a value-maximizing multi-hop cooperation path for the current task through their agents in a distributed manner, enabling efficient and reliable multi-device collaboration. The main contributions of this paper are summarized as follows.
\begin{itemize}[leftmargin = *]
   \item We propose the GADAI framework to address the selection of task-specific trusted multi-hop collaborators. To enhance the accuracy of task-specific trust evaluation, we design an effective mechanism that performs independent evaluation of each trust-related perspective and integrates the results for decision-making.

   \item We develop the GNN-assisted model to infer the historical collaboration reliability of devices. The model effectively integrates trust information from multi-hop neighbours, thereby achieving more robust and accurate assessments of devices’ historical reliability.

   \item We innovatively utilize agentic AI to enable devices to autonomously perform task-specific trust evaluation, effectively addressing the issues of resource privacy protection and dynamic change. 

    \item We propose an LAM-enabled multi-agent collaboration approach that enables devices in distributed systems to collaboratively plan task-oriented multi-hop paths, enhancing system cooperation efficiency.
\end{itemize}

The rest of the paper is organized as follows. A comprehensive review of the literature is provided in Section~\ref{section_work}. System model and problem formulation are introduced in
Section~\ref{section_system_model}. The GNN-aided evaluation model
is presented in Section~\ref{section_GNN}. Section~\ref{section_GAI} introduces the use of agentic AI for resource trust evaluation and multi-hop path planning. Simulation results are provided in
Section~\ref{section_experiment}. Finally, Section~\ref{section_conclusion} concludes this paper.

\begin{table*}[!t] 
        \footnotesize  
	\centering
        \renewcommand{\arraystretch}{1.3}
	\caption{A comparison of six dimensions between our work and existing research.}
	\label{related_comparison}
	  \begin{tabular}{m{3cm}<{\centering}|m{2cm}<{\centering}|m{1.5cm}<{\centering}|m{2cm}<{\centering}|m{1.5cm}<{\centering}|m{2.5cm}<{\centering}|m{2.5cm}<{\centering}}
		\hline {Reference} & {Historical reliability evaluation} & {Task attribute} & {Resource evaluation}  & {Resource privacy} & {Task-specific resource evaluation} & {Multi-hop collaboration}  \\
		\hline \hline
            \cite{8769947}, \cite{10103199}, \cite{10104094}, \cite{9794601}, \cite{6519238}  & $\surd$  &  $\times$ & $\times$ & $\times$ &  $\times$ & $\times$\\
             \hline
             \cite{9305298}, \cite{10320384}, \cite{10251781}  & $\surd$  & $\times$  & $\surd$ & $\times$ &  $\times$ & $\times$ \\
             \hline
             \cite{10812999}, \cite{10577995}, \cite{10239502}, \cite{9795858}  &  $\times$ &  $\times$ & $\times$  &  $\times$ & $\times$  & $\surd$ \\
             \hline
             \cite{9335844}, \cite{7986943}, \cite{doi:10.1155/2014/209436}  &  $\surd$ &  $\times$& $\times$ & $\times$ & $\times$ & $\surd$\\
             \hline
             \cite{7194748} & $\surd$ & $\times$ &  $\surd$ & $\times$ & $\times$ & $\surd$ \\
             \hline
             Our work & $\surd$ & $\surd$ & $\surd$ & $\surd$ & $\surd$ & $\surd$ \\
             \hline
             \multicolumn{6}{l}{Note: $\surd$ indicates this factor is considered, and $\times$ indicates it is not.} \\
             \hline
   \end{tabular}
\end{table*}

\section{Related Work}
\label{section_work}
In this section, we first present an in-depth review of existing research on trust-based collaborator selection.  Next, we investigate multi-hop path planning. A comparison of our work with existing research is provided in Table~\ref{related_comparison}.

\vspace{-0.08 in}
\subsection{Trust-Based Collaborator Selection}
Trust, as an effective tool for evaluating the reliability of collaborators, is gaining increasing attention in collaborative systems. In~\cite{10422726}, the authors proposed a dynamic long-term and short-term trust calculation model based on feedback to identify honest devices in the industrial internet. The authors in~\cite{10103199} assessed trust based on task completion feedback and used it to identify reliable collaborators within the Internet of Vehicles. In~\cite{10104094}, the authors identified trusted collaborators by evaluating their behaviour in executing virtual reality video streaming tasks in a mobile edge computing system. In~\cite{9794601}, a semi-centralized trust management system was proposed that calculates trust based on both direct experiences and indirect recommendations. In~\cite{8769947}, the authors assessed the trustworthiness of sensor nodes based on their data collection and communication behaviour. In~\cite{6519238}, the authors evaluated the trustworthiness of devices in heterogeneous wireless networks by collecting multidimensional interaction information, such as packet forwarding success rate and session interruption rate. The studies mentioned above primarily use feedback, recommendations, and historical interactions to evaluate device trust, neglecting that device resources should also be a key component of trust.

Recently, some researchers have started incorporating device resources, such as communication and computation, into trust evaluation to more accurately reflect device trustworthiness. In~\cite{10320384}, the authors proposed a fast and reliable trust evaluation method, utilizing historical experience and real-time resource conditions to select collaborators in the Internet of Vehicles. The authors in~\cite{10251781} evaluated the trust levels of edge computing servers according to their bandwidth, computing ability, and initial success rate. In~\cite{9305298}, device trust was evaluated by integrating computing ability, social relationships, external opinions, and dynamic knowledge. However, these methods fail to account for two critical aspects. First, device resource information may be unsuitable for public sharing due to privacy concerns. Second, they neglect the influence of task attributes on the trustworthiness of device resources, leading to resource evaluations that are not task-specific. 

Different from them, we comprehensively integrate two evaluation dimensions—historical reliability assessment and resource trust evaluation—to achieve accurate assessment results. In particular, task attributes and the privacy protection of resource information are carefully considered.

\vspace{-0.08 in}
\subsection{Multi-Hop Path Planning}
In collaborative systems, efficiently planning multi-hop cooperative paths to connect distant nodes is an urgent challenge. Some researchers have begun exploring various methods to address this issue in wireless systems. In~\cite{9795858}, the authors used convolutional neural networks and reinforcement learning to develop a packet forwarding path algorithm for minimizing transmission time in a multi-hop UAV relay network. In~\cite{10239502}, the authors proposed a multi-agent deep deterministic policy gradient method to solve the task routing problem in the vehicle-assisted collaborative edge computing system.  The authors in~\cite{10812999} introduced the improved Q-learning-based multi-hop routing algorithm that facilitates energy-efficient and reliable data transmission in UAV-assisted communication. 
To solve the multi-hop task offloading problem in the Internet of Vehicles, the authors in~\cite{10577995} proposed an asynchronous deep reinforcement learning approach to optimize the selection of forwarding vehicles, aiming to minimize task delay. However, these studies lack mechanisms to guarantee the trustworthiness of selected multi-hop collaborators. In addition, their machine learning approaches demand large amounts of training data, rendering them unsuitable for dynamic and constantly evolving collaborative systems. 

 \begin{table}[!t] 
	\centering
        \renewcommand{\arraystretch}{1.2}
	\caption{Summary of Notations}
	\label{notation}
	  \begin{tabular}{m{1.1cm}<{}|m{6.8cm}<{}}
		\hline \textbf{Notation} & \textbf{Description} \\
		\hline \hline
              $A$ & The set of terminal/task forwarding (TF) devices, \{$a_1$,\dots, $a_I$\}\\  
              \hline
              $B$ & The set of edge computing (EC) devices, \{$b_1$, \dots, $b_M$\}\\
              \hline
              
               $c^{\text{TF}}$ & The minimum trust threshold of task $C$ for TF devices \\
               \hline
               $c^{\text{EC}}$ &  The minimum trust threshold of task $C$ for EC devices\\
              \hline
               $D$ & The set of historical collaborations, $d_{(a_i,b_m)}, d_{(a_i,a_j)} \in D$\\ 
              \hline
              $G^{\text{top}}$ & Network topology\\
              \hline
              $G^{\text{new}}$ & A network topology that only includes trusted devices based on historical collaborations \\
            \hline
            $G^{\text{dir}}$ &  Historical collaboration graph\\
            \hline
              $h^{\text{tr}}_{a_j}$ & The embedding of device $a_j$ in the $l$-th GNN layer when acting as the trustor \\
              \hline
              $h^{\text{te}}_{a_j}$ & The embedding of device $a_j$ in the $l$-th GNN layer when acting as the trustee \\
              \hline
              $h^{(l)}_{a_j}$ & The comprehensive embedding of device $a_j$ in the $l$-th GNN layer \\
              \hline
              $N_{(a_i,a_j)}$ & The total number of tasks received by device $a_j$ from device $a_i$ \\
    \hline
    $s^{\text{TF}}_{\text{soft}}$ & The soft threshold of the fee the task owner is willing to pay to a TF device \\
               \hline
               $s^{\text{TF}}_{\text{hard}}$ & The hard threshold of the fee the task owner is willing to pay to a TF device\\
              \hline
               $s^{\text{EC}}_{\text{soft}}$ &  The soft threshold of the fee the task owner is willing to pay to an EC device\\
               \hline
               $s^{\text{EC}}_{\text{hard}}$ & The hard threshold of the fee the task owner is willing to pay to an EC device\\
               \hline
              $T_{(a_i,a_j)}$ & Task forwarding trust of device $a_j$ evaluated by device $a_i$ \\
              \hline
              $T_{(a_i,a_j)}^{\text{his}}$ & Historical reliability of device $a_j$ evaluated by device $a_i$ \\
              \hline
              $T_{(a_i,a_j)}^{\text{res}}$ & Trustworthiness of device $a_j$'s resource evaluated by device $a_i$ \\
              \hline
               $T_{(a_i,b_m)}$ & Task computing trust of device $b_m$ evaluated by device $a_i$ \\
              \hline
              $T_{(a_i,b_m)}^{\text{his}}$ & Historical reliability of device $b_m$ evaluated by device $a_i$\\
              \hline
              $T_{(a_i,b_m)}^{\text{res}}$ & Trustworthiness of device $b_m$'s resource evaluated by device $a_i$ \\
              \hline
                VoC & Value of task completion \\
              \hline
              $\overline{V}_{\pi(a_i,b_m)}$ & Average VoC perceived by the task owner from devices on the path  $\pi_{(a_i,b_m)}$\\
              \hline
               $\overline{V}^{\text{max}}_{(a_i,b_j)}$ & Maximum average VoC perceived by the task owner from devices on the path between device $a_i$ and device $a_j$\\
               \hline
                $\pi_{(a_i,b_m)}$ & A trusted multi-hop collaboration path from the task owner to a potential EC device $b_m$\\ 
              \hline
   \end{tabular}
\end{table}

In recent years, some researchers have investigated incorporating security or trust mechanisms into the selection of multi-hop collaborators. The authors in \cite{7986943} proposed a security-oriented routing mechanism specifically designed for Software-Defined Networking (SDN), which leverages both the network performance and security capabilities of SDN switch nodes to improve overall network reliability and security. In \cite{doi:10.1155/2014/209436}, the authors integrated trust evaluations into the routing decisions, allowing nodes to assess the reliability of their neighbours based on past interactions. In \cite{7194748}, trust evaluation and energy awareness are incorporated into the routing to address challenges such as malicious node behaviour and limited energy resources. In \cite{9335844}, the authors modified the traditional Dijkstra’s algorithm for multi-hop relay networks to account for trust degrees, allowing paths to be chosen that both minimize transmission time and meet a minimum cumulative trust threshold, thus avoiding untrusted nodes. Although these methods incorporate trust mechanisms into multi-hop routing algorithms, they provide only general trust-aware multi-hop frameworks without task-specific considerations, resulting in limited adaptability to dynamic environments.

Recent studies, such as \cite{vanstein2024llamea}, \cite{NEURIPS2024_4ced59d4}, and \cite{Liu2024EvolutionOH}, have leveraged the reasoning and adaptive capabilities of LAMs to tackle multi-hop path planning challenges in combinatorial optimization. These approaches integrate LAMs with evolutionary computation to automatically generate or refine heuristics, achieving promising results in problems such as the travelling salesman problem and the capacitated vehicle routing problem. Inspired by recent work on LAMs and the challenges of trusted multi-hop path planning, our study employs the LAM-enabled agentic AI system to perform autonomous, task-specific, and distributed multi-hop path planning in dynamic environments.

\section{System Model and Problem Formulation}
\label{section_system_model}

\begin{figure}[t!]
\centering
\includegraphics[scale=0.95]{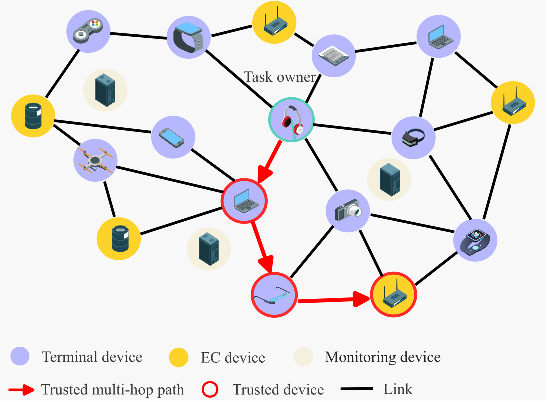}
\caption{A task is transmitted from the task owner through a sequence of trusted terminal devices to a trusted edge computing device.}
\label{systemmodel}
\end{figure}

A collaborative system is considered, consisting of a set of terminal devices $A=\{a_1,\dots,a_I \}$ and a set of edge computing (EC) devices $B=\{b_1,\dots, b_M \}$. Terminal devices, such as mobile phones, can function either as task owners that generate tasks or as task forwarding (TF) devices that relay tasks for other terminals. EC devices, equipped with computational resources, offer computing services for terminal devices. All terminal and EC devices form a network topology denoted as $G^{\text{top}} = ((A, B), {E}^{\text{top}})$, where ${E}^{\text{top}}$ is the set of communication links among devices. As shown in Fig.~\ref{systemmodel}, due to geographical constraints, a task generated by a task owner must be relayed through multiple trusted terminal devices before reaching a trusted EC device. The system also includes a set of monitoring devices that collect collaboration data. Communication between monitoring devices is conducted through an independent system. The following subsections provide detailed descriptions of the task model, trusted multi-hop collaboration model, task forwarding model, and task computing model. The main notations used in this paper are summarized in Table~\ref{notation}.

\vspace{-0.09 in}
\subsection{Task Model}
A task $C$,  initiated by a terminal device $a_i$, acting as a task owner, is parameterized as ($c^{\text{des}}$, $c^{\text{size}}$, $c^{\text{TF}}$, $c^{\text{EC}}$, $s^{\text{TF}}_{\text{soft}}$, $s^{\text{TF}}_{\text{hard}}$, $s^{\text{EC}}_{\text{soft}}$, $s^{\text{EC}}_{\text{hard}}$), where $c^{\text{des}}$ represents the processing density (cycles/bit), $c^{\text{size}}$ denotes the number of data bits, $c^{\text{TF}}$ and $c^{\text{EC}}$ are the minimum trust thresholds for terminal devices and EC devices, respectively. Furthermore, $s^{\text{TF}}_{\text{soft}}$ and $s^{\text{TF}}_{\text{hard}}$ defines the soft and hard fee thresholds that the task owner is willing to pay a terminal device, while $s^{\text{EC}}_{\text{soft}}$ and $s^{\text{EC}}_{\text{hard}}$ serve as the same purpose for an EC device. These parameters are determined by the task owner's requirements.

\vspace{-0.06 in}
\subsection{Task-Specific Trusted Multi-Hop Collaboration Model}
To ensure reliable collaboration, all devices involved in task forwarding or task computing must satisfy the trust thresholds specified by task $C$. Hence, accurate trust evaluation for terminal and EC devices is essential. The trust evaluation models for terminal and EC devices are defined as follows.

\begin{Definition}[Task forwarding trust]
    \textnormal{The task forwarding trust that the task owner $a_i$ places in a terminal device $a_j$ is defined as $a_i$'s expectation of $a_j$'s ability to successfully forward task $C$. This trust is formally calculated as:
    \vspace{-0.05 in}
    \begin{equation}
        \label{final_a_i_a_j}
        T_{(a_i,a_j)} = T_{(a_i,a_j)}^{\text{his}} T^{\text{res}}_{(a_i,a_j)}, 
    \end{equation}
    where $T_{(a_i,a_j)}^{\text{his}} \in [0,1]$ represents the historical collaboration reliability of device $a_j$, and $T^{\text{res}}_{(a_i,a_j)} \in [0,1]$ reflects the resource trust of device $a_j$ specific to task $C$.}
\end{Definition} 
\begin{Definition}[Task computing trust]
\textnormal{The task computing trust that the task owner $a_i$ places in an EC device $b_m$ is defined as $a_i$'s expectation of $b_m$'s ability to execute task $C$. This trust is formally calculated as:
   \vspace{-0.1 in}
    \begin{equation}
        \label{final_a_i_b_m}
        T_{(a_i, b_m)} = T_{(a_i, b_m)}^{\text{his}} T^{\text{res}}_{(a_i, b_m)},  
    \end{equation}
where $T_{(a_i,b_m)}^{\text{his}} \in [0,1]$ denotes the reliability of device $b_m$ based on historical collaboration data $D$, while $T^{\text{res}}_{(a_i,b_m)} \in [0,1]$ represents its resource trust with respect to task $C$.}
\end{Definition}

As indicated in Definitions 1 and 2, accurate evaluation requires a comprehensive assessment of both devices’ historical collaboration data and their current resource availability. The historical collaboration datasets $D$ are collected by monitoring devices.
We assume that a collaboration $d_{(a_i,a_j)} \in D$ represents an instance where device $a_j$ assists in forwarding a task from the task owner $a_i$, recording the relevant performance of device $a_j$ during this activity. Similarly, a collaboration $d_{(a_i,b_m)} \in D$ denotes the case where an EC device $b_m$ executes a task generated by the task owner $a_i$, capturing $b_m$'s performance. Given that task $C$ must be relayed through multiple hops to reach an EC device $b_m$, a multi‑hop collaboration path is established between the task owner $a_i$ and $b_m$. Based on the network topology $G^{\text{top}}$, the set of all possible paths from the task owner $a_i$ to $M$ EC devices is denoted by ${\Phi}$. A path $\pi_{(a_i,b_m)} \in \Phi$ is regarded as a task-specific trusted multi‑hop collaboration path if and only if every device on the $\pi_{(a_i,b_m)}$ satisfies the trust thresholds of task $C$. The path $\pi_{(a_i,b_m)}$ is formally defined as follows.

\begin{Definition}[Task-specific trusted multi-hop collaboration path]\textnormal{
   A path from the task owner $a_i$ to an EC device $b_m$ via a $K$-hop route is defined as a task-specific trusted multi-hop path if the trustworthiness of every terminal device $a_j$ along the path satisfies $T_{(a_i,a_j)} \geq c^{\text{TF}}$ and the trustworthiness of the EC device $b_m$ satisfies $T_{(a_i,b_m)} \geq c^{\text{EC}}$. Formally, it is represented as:
    \begin{align}
        f^{a_i}_0 a_i \rightarrow\cdots f^{a_r}_{k-1} a_r \to f^{a_j}_{k}a_j \rightarrow f^{a_p}_{k+1}a_p\cdots \rightarrow f^{b_m}_K b_m,
    \end{align}
    where $f^{a_j}_k$ is a binary variable that equals 1 if device $a_j$ occupies the $k$-th hop on the path, and 0 otherwise.}
\end{Definition}

\vspace{-0.05 in}
\subsection{Task Forwarding and Task Computing Models}

Task $C$ is relayed sequentially along the terminal devices on the path $\pi_{(a_i,b_m)}$ until it reaches the EC device $b_m$. The task forwarding and computing models are detailed below.

\textit{1) Task forwarding model}:
All terminal devices are assumed to operate in a receive-and-forward mode: each device first receives task $C$, stores it in a local buffer, and then forwards it to the next device. The buffer size of each device is considered equal to the current available storage capacity. The time required by the $k$-th hop device $a_j$ to receive task $C$ from the $(k-1)$-th hop device $a_r$ is given by:
\begin{align}
  \label{t_rec}
    t_{(k-1,k)} = \frac{c^{\text{size}}}{\delta_{(k-1,k)}},
\end{align}
where $\delta_{(k-1,k)}$ denotes the average transmission rate from the $(k-1)$-th hop device $a_r$ to the $k$-th hop device $a_j$. It is calculated as:
\begin{align}
    \delta_{(k-1,k)} = W^{\text{band}} \log_2 \left( 1 + \frac{a_{r}^{\text{pow}}h_{(a_r,a_j)}}{N_0} \right),
\end{align}
where $W^{\text{band}}$ denotes the bandwidth, $a^{\text{pow}}_r$ represents the transmission power of device $a_r$, and $N_0$ is the noise power. The channel gain is given by $h_{(a_r,a_j)} = |a_r - a_j|^{-4}$, where $|a_r - a_j|$ denotes the distance between $a_r$ and $a_j$~\cite{7307234}. Similarly, the time required by device $a_j$ to send task $C$ to the $(k+1)$-th hop device $a_p$ can be calculated as $t_{(k, k+1)}$. As device $a_j$ charges for the task forwarding service, the fee paid by the task owner $a_i$ is calculated using a time-based pricing model, which covers both the receiving and sending phases:
\begin{align}
    s_{k} = a_j^{\text{pri}}\left(t_{(k-1,k)} + t_{(k,k+1)} \right),
\end{align}
where $a_j^{\text{pri}}$ represents $a_j$'s service price $(\$/\text{second})$. To assess the task owner’s perceived satisfaction with task forwarding, it is essential to define an appropriate evaluation metric. The concept of value is widely used in marketing activities and is usually used to represent the benefits users get from services or products. Inspired by the approaches in \cite{9447211} and \cite{10.1145/2913712.2913716}, where perceived value is employed as a metric for assessing service quality, this study adopts perceived value to evaluate the quality of task forwarding from the perspective of the task owner. Accordingly, the value of task completion (VoC) perceived by the task owner $a_i$ for the task forwarding service provided by the $k$-th hop device $a_j$ is calculated as:
    \begin{align}
    \label{votf}
   \hspace{-0.1in}V_{k} = \begin{cases}
    1, & \text{if} \  s_{k} < s^{\text{TF}}_{\text{soft}}; \\
    e^{-\left|\frac{s_{k} - s^{\text{TF}}_{\text{soft}}}{s^{\text{TF}}_{\text{soft}}}\right|}, & \text{if} \   s^{\text{TF}}_{\text{soft}} \le s_{k} \le s^{\text{TF}}_{\text{hard}}; \\
    0, & s_{k} > s^{\text{TF}}_{\text{hard}},
   \end{cases}
   \end{align} 
where $s^{\text{TF}}_{\text{soft}}$ and $s^{\text{TF}}_{\text{hard}}$ define the soft and hard fee thresholds, respectively. The perceived VoC reaches its maximum value of 1 when the actual fee $s_k$ is below $s^{\text{TF}}_{\text{soft}}$, and drops to 0 when $s_k$ exceeds $s^{\text{TF}}_{\text{hard}}$. When the actual fee falls between these thresholds, the VoC is determined by comparing the actual fee with the soft threshold using the exponential decay function derived from Kano's quality theory, which is commonly used to model user satisfaction~\cite{7852434}.

\textit{2) Task computing model}:
After receiving task $C$ from the $(K-1)$-th hop device, the last-hop EC device on the path $\pi_{(a_i,b_m)}$ executes the task computing. The computing time is given by~\cite{10546264}:
\vspace{-0.08 in}
\begin{align}
   \label{processing_time}
    t_{(K-1,K)} = \frac{c^{\text{des}}c^{\text{size}}}{b^{\text{cpu}}_{m}}, 
\end{align}
where $b^{\text{CPU}}_{m}$ is the available CPU clock frequency of device $b_m$. The fee charged by device $b_m$ for this computing service is calculated as $s_{K} = t_{(K-1,K)} b^{\text{pri}}_m$, where $ b^{\text{pri}}_m$ is the service price ($\$/\text{second}$). Similarly, the VoC perceived by the task owner during this process is calculated as:
\vspace{-0.05 in}
    \begin{align}
     \hspace{-0.1in} V_{K} = \begin{cases}
    1, & \text{if} \  s_{K} < s^{\text{EC}}_{\text{soft}}; \\
    e^{-\left|\frac{s_{K} - s^{\text{EC}}_{\text{soft}}}{s^{\text{EC}}_{\text{soft}}} \right|}, & \text{if} \   s^{\text{EC}}_{\text{soft}} \le s_{K} \le s^{\text{EC}}_{\text{hard}}; \\
    0, & s_{K} > s^{\text{EC}}_{\text{hard}},
   \end{cases}
   \end{align}
where $s^{\text{EC}}_{\text{soft}}$ and $s^{\text{EC}}_{\text{hard}}$ are the soft and hard thresholds for the fee that the task owner is willing to pay the EC device $b_m$. After all $K$ devices on the path $\pi_{(a_i,b_m)}$ complete the forwarding and computing of task $C$, the average VoC perceived by the task owner $a_i$ is calculated as:
\vspace{-0.06 in}
\begin{align}
     \label{average_value}
    \overline{V}_{\pi_{(a_i,b_m)}} = \sum_{k=1}^{K} V_{k}/K. 
\end{align}

\vspace{-0.2 in}
\subsection{Problem Formulation}
 As we can see, the average VoC is determined by the selection of trusted terminal devices and a trusted EC device, and an efficiently planned path connecting them. To maximize the average VoC, this research formulates a joint optimization problem of trusted device selection and path planning: 
\begin{alignat}{1}
     \label{problem}
     &\max_{{\Phi}, {A}, {B}} \quad \overline{V}_{\pi_{(a_i,b_m)}},   \\
    \mathrm{s.t.} \quad
    & \pi_{(a_i,b_m)} \in \Phi, \label{sub-pi} \\
    &  T_{(a_i,a_j)} \ge c^{\text{TF}}, \forall a_j \in \pi_{(a_i,b_m)} \, \text{and} \, A, \label{sub-1}\\
    &  T_{(a_i,b_m)} \ge c^{\text{EC}},  b_m \in \pi_{(a_i,b_m)} \, \text{and} \, B, \label{sub-2}\\
    & \sum_{k=1}^{K-1} f^{a_j}_k = 1,  \quad \forall a_j \in \pi_{(a_i,b_m)} \label{sub-3}\\ 
    & f^{a_i}_0 = 1, f^{b_m}_K = 1.   \label{sub-4}
\end{alignat}
  Constraint (\ref{sub-pi}) restricts the scope of path planning to the set of all possible paths $\Phi$ in $G^{\text{top}}$. Constraint (\ref{sub-1}) states that the trustworthiness of all selected terminal devices should meet the minimum trust threshold $c^{\text{TF}}$. Constraint (\ref{sub-2}) specifies that the trustworthiness of the selected EC device must meet the minimum trust threshold $c^{\text{EC}}$. Constraint (\ref{sub-3}) stipulates that each selected terminal device can appear at only one position on the path $\pi_{(a_i,b_m)}$, preventing multiple visits. 
 Constraint (\ref{sub-4}) states that the planned path $\pi_{(a_i,b_m)}$ starts at the task owner and terminates at the selected EC device $b_m$.

\section{GNN-Aided Historical Reliability Evaluation}
\label{section_GNN}

To achieve multi-hop path planning that maximizes the average VoC, we propose the GADAI framework. It first conducts a two-stage trust evaluation for devices: utilizing the GNN-aided model for historical reliability assessment, and employing the agentic AI system to perform task-specific resource trust evaluation. The evaluation results are then integrated to identify trusted collaborators. Through their agents, these trusted collaborators collectively construct a task-oriented multi-hop cooperation path in a distributed manner. In this section, we present the GNN-aided evaluation model, which involves four steps: generating the historical collaboration graph, propagating and aggregating trust information, calculating historical reliability, and removing devices that fall below the trust thresholds.

\begin{figure*}[t!]
\centering
\includegraphics[scale=0.99]{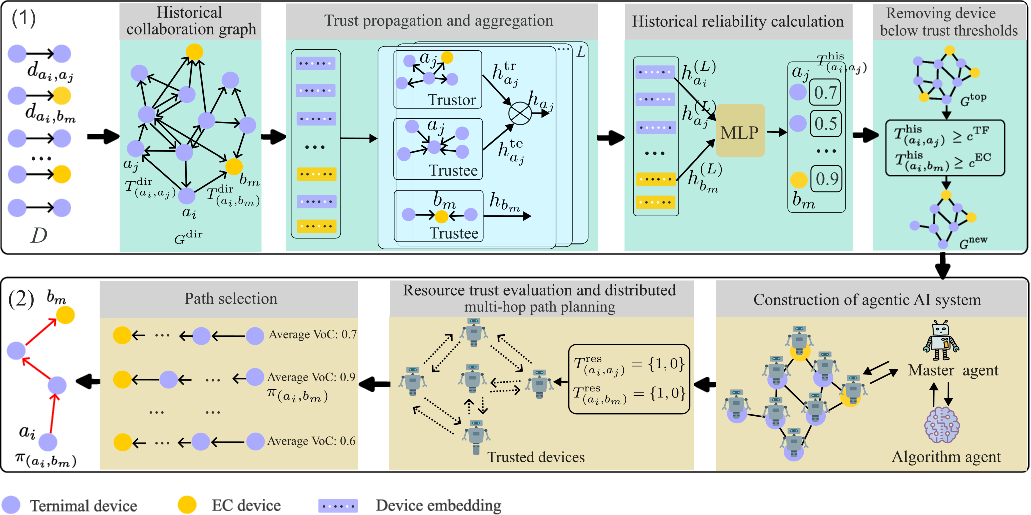}
\caption{The proposed GADAI framework. (1) Historical reliability evaluation of devices involves four steps: historical collaboration graph construction, trust propagation and aggregation, historical reliability computation, and the removal of devices below the trust thresholds. (2)~Task-specific device resource trust assessment and multi-hop path planning enabled by the agentic AI system.}
\label{gnn}
\end{figure*}

\vspace{-0.06 in}
\subsection{Historical Collaboration Graph Construction}
According to the collected historical collaboration dataset $D$ between devices, their direct trust relationships can be mined. To comprehensively assess the capabilities of terminal devices, we use the packet loss rate (PLR) to evaluate the quality of communication links and the task forwarding success rate (TFSR) to measure their forwarding capability. Hence, each collaboration $d_{(a_i,a_j)}$ includes device $a_j$'s PLR and TFSR. Accordingly, an edge $e_{(a_i,a_j)}$ is established from device $a_i$ to device $a_j$. The weight of $e_{(a_i,a_j)}$ represents the direct trust $T_{(a_i, a_j)}^{\text{dir}}$ that device $a_i$ places in device $a_j$, which is computed from all collaborations from device $a_i$ to device $a_j$:
\begin{align}
    \label{a_i_dir_a_j}
   \hspace{-0.098 in} T_{(a_i,a_j)}^{\text{dir}} =  \frac{1}{N_{(a_i, a_j)}}\sum_{n=1}^{N_{(a_i,a_j)}}\left( \alpha_1 \left(1 - T_{n}^{\text{PLR}}\right) +  \alpha_2 T_{n}^{\text{TFSR}} \right), 
\end{align}
where $N_{(a_i,a_j)}$ is the total number of tasks received by device $a_j$ from device $a_i$, $\alpha_1$ and $\alpha_2$ are weight coefficients, $0 \leq \alpha_1, \alpha_2 \leq 1$, $ \alpha_1 + \alpha_2 = 1$. $T_{n}^{\text{PLR}}$ and $T_{n}^{\text{TFSR}}$ measure the communication link quality from device $a_i$ to device $a_j$ and the packet forwarding ability of $a_j$, respectively. These metrics are calculated as: 
\begin{align}
    T_{n}^{\text{PLR}} = \frac{P^{\text{lost}}_n}{P^{\text{tot}}_n},\, T_{n}^{\text{TFSR}} = \frac{P_n^{\text{tra}}}{P_n^{\text{rec}}},
\end{align}
where $P^{\text{tot}}_n$ denotes the total number of packets sent from device $a_i$ to device $a_j$ in the $n$-th task, and $P^{\text{lost}}_{n}$ indicates the total number of packets lost during transmission within the same task. $P_{n}^{\text{rec}}$ is the total number of received packets by device $a_j$ from device $a_i$ in the $n$-th task, while $P_{n}^{\text{tra}}$ represents the number of packets successfully forwarded by device $a_j$.

For EC devices, task outcomes are collected to evaluate their performance. A collaboration $d_{(a_i,b_m)}$ records the result of a task executed by device $b_m$ from device $a_i$, where a successful outcome is denoted by 1 and a failure by 0. Accordingly, an edge $e_{(a_i, b_m)}$ is established from device $a_i$ to device $b_m$. The weight of $e_{(a_i, b_m)}$ represents the direct trust $T^{\text{dir}}_{(a_i,b_m)}$ that device $a_i$ places in device $b_m$, which is calculated as follows:
\begin{align}
 \label{a_i_dir_b_m}
    T^{\text{dir}}_{(a_i,b_m)} = \frac{\sum^{N_{(a_i,b_m)}}d_{(a_i,b_m)}}{N_{(a_i,b_m)}},
\end{align}
where $N_{(a_i,b_m)}$ is the total number of tasks executed by device $b_m$ from device $a_i$. From the historical dataset $D$, we ultimately obtain a historical collaboration graph that represents the direct trust relationships among devices, denoted as $G^{\text{dir}} = ((A, B), {E}^{\text{dir}}, {W}^{\text{dir}}, N^{\text{dir}})$, where ${E}^{\text{dir}}$ is the set of all edges, ${W}^{\text{dir}} = \{\dots T^{\text{dir}}_{(a_i,a_j)}, T^{\text{dir}}_{(a_i,b_m)} \dots \}$ is the set of edge weights, and $N^{\text{dir}} = \{\dots, N_{(a_i, a_j)}, N_{(a_i,b_m)}, \dots\}$ is the set of collaboration frequencies among devices.

\subsection{Propagation and Aggregation of Trust Information} 

To accurately evaluate the trustworthiness of devices, trust information needs to be propagated and aggregated among them. The propagation process transfers device trust information across multi-hop neighbours, enabling the dissemination of trust influence beyond direct collaborations. The aggregation process integrates trust information from multiple neighbours to generate an accurate and robust reliability evaluation for each device. Therefore, we first perform the embedding operation for all devices, and then use the GNN model to propagate and aggregate trust information among them, as presented in Algorithm~\ref{gnn-trust-framework}.  

Embedding layer: Initial device embeddings are generated using node2vec~\cite{10.1145/2939672.2939754}, where each terminal device $a_i$ and each EC device $b_m$ is mapped into an $H_a$-dimensional vector space, yielding embeddings $h_{a_i}$ and $h_{b_m}$.

 Trust propagation and aggregation layer: Given that terminal devices participate in task workflows by both
receiving and forwarding tasks, they inherently act as
trustors and trustees. Accordingly, their embeddings should
effectively reflect this dual functionality. Specifically, within the historical collaboration graph $G^{\text{dir}}$, a device’s out-degree represents the collaborations that it initiates (acting as a trustor), while its in-degree represents the collaborations that it receives (acting as a trustee). The trust propagation and aggregation mechanism for these dual roles is presented below.
    
{\textit{1) First-Order Trust Propagation and Aggregation}

\textbf{Trust propagation and aggregation when terminal devices act as trustees}: When a terminal device $a_j$ acts as a trustee, its first-order in-degree neighbours in the graph propagate their trust values to device $a_j$. This process is formulated as:
\begin{align}
    \label{w_a_j_get_a_i}
    \omega_{a_j \gets a_i} &= W_{a_j \gets a_i} \cdot \chi_{a_j \gets a_i}, \\
    \label{ep_a_j_get_a_i}
    \epsilon_{a_j \gets a_i} &= W_{a_j \Leftarrow a_i} \cdot \eta_{a_j \gets a_i},\\ 
    \label{mu_a_j_get_a_i}
    \mu_{a_j \gets a_i}  &= h_{a_i} \parallel \omega_{a_j \gets a_i} \parallel \epsilon_{a_j \gets a_i},
\end{align}
where $ \chi_{a_j \gets a_i} \in \mathbb{R}^{H_T \times 1}$ is the embedding of trust value $T^{\text{dir}}_{(a_i,a_j)}$ that $a_i$ places in $a_j$, which is transformed by using binary encoding~\cite{9945661}. $\eta_{a_j \gets a_i} \in \mathbb{R}^{H_T \times 1}$ is the embedding of the normalized collaboration frequency $N_{(a_i,a_j)}$. $W_{a_j \Leftarrow a_i}, W_{a_j \gets a_i} \in \mathbb{R}^{H_a \times H_T}$ are the trainable transformation matrices, and $\parallel$ is a concatenation operation. $\mu_{a_j \gets a_i}$ can be interpreted as device $a_i$'s recommendation for device $a_j$ based on $N_{(a_i,a_j)}$ collaborations. After receiving the messages from its first-order neighbours, device $a_j$ aggregates them. Although mean aggregation is commonly adopted in previous studies, it fails to distinguish the varying importance of different neighbours. Since frequently and infrequently cooperating neighbours have different levels of importance, an effective model should be capable of assigning different weights to distinguish their relative significance. Therefore, we apply an attention mechanism to calculate the importance of neighbours. Following \cite{9488814}, the importance of each in-degree neighbour of device $a_j$ is calculated as the inner product of the message from device $a_i$ to device $a_j$ and the embedding of device $a_j$, as expressed by: 
   \begin{align}
    \bar{\psi}_{a_j\gets a_i} &= \text{attention} (\mu_{a_j \gets a_i}W_{\mu_{a_j \gets a_i}}, h_{a_j}W_{h_{a_j}}), \nonumber\\
    &= \mu_{a_j \gets a_i}W_{\mu_{a_j \gets a_i}} \cdot (h_{a_j}W_{h_{a_j}})^{\top},
   \end{align} 
where $W_{\mu_{a_j \gets a_i}} \in \mathbb{R}^{(H_a + H_T + H_T) \times H_a}$ and $W_{h_{a_j}} \in \mathbb{R}^{H_a \times H_a}$ are the trainable matrices. Then, we normalize $ \bar{\psi}_{a_j\gets a_i}$ using the \textit{softmax} function:
\begin{align}
    {\psi}_{a_j\gets a_i} = \frac{\text{exp}(\bar{\psi}_{a_j\gets a_i})}{\sum_{a_i \in {\mathcal{N}}^{\text{in}}_{a_j}}\text{exp}(\bar{\psi}_{a_j\gets a_i})},
\end{align}
where ${\mathcal{N}}^{\text{in}}_{a_j}$ is the set of in-degree neighbors of device $a_j$. 
Based on the single-head attention mechanism, a multi-head attention fusion is further applied to enhance representation capacity. Accordingly, device $a_j$ aggregates messages from its in-degree neighbours through the multi-head attention fusion as follows:
\begin{align}
    \label{h_a_j_te}
    h_{a_j}^{\text{te}} &= \text{Aggregate}\left(\mu_{a_j \gets a_i}, a_i \in \mathcal{N}^{\text{in}}_{a_j} \right), \nonumber\\ 
    &= \parallel_{Q} \sigma \left(\left(\sum_{a_i \in \mathcal{N}^{\text{in}}_{a_j}} {\psi}_{a_j\gets a_i} \mu_{a_j \gets a_i} \right) \cdot W_{\text{agg}} \right),
\end{align}
where $Q$ is the number of heads, and $\sigma$ is a non-linear activation function~\cite{ECKLE2019232}.

\textbf{Trust propagation and aggregation when terminal devices act as trustors}: 
When acting as a trustor, device $a_j$ receives trust and cooperation frequency information from its out-degree neighbours, which is then aggregated through a multi-head attention mechanism:
\begin{align}
    \label{omega_a_j_to_a_i}
    \omega_{a_j \to a_i} &= W_{a_j \to a_i} \cdot \chi_{a_j \to a_i}, \\
    \label{ep_a_j_to_a_i}
    \epsilon_{a_j \to a_i} &= W_{a_j \Rightarrow a_i} \cdot \eta_{a_j \to a_i}, \\
    \label{mu_a_j_to_a_i}
    \mu_{a_j \to a_i}  &= h_{a_i} \parallel \omega_{a_j \to a_i} \parallel \epsilon_{a_j \to a_i},\\
    \label{h_a_j_tr}
    h_{a_j}^{\text{tr}} &= \text{Aggregate} \left(\mu_{a_j \to a_i}, a_i \in {\mathcal{N}}^{\text{out}}_{a_j} \right),
\end{align}
where $W_{a_j \Rightarrow a_i}, W_{a_j \to a_i} \in \mathbb{R}^{H_a \times H_T}$ are the learnable transformation matrices, $\chi_{a_j \to a_i}$ is the embedding of direct trust value $T^{\text{dir}}_{(a_j,a_i)}$ that $a_j$ places in $a_i$, $\eta_{a_j \to a_i}$ is the embedding of the normalized collaboration frequency $N_{(a_j,a_i)}$, 
and ${\mathcal{N}}^{\text{out}}_{a_j}$ is the set of out-degree neighbors of device $a_j$. 

\begin{algorithm}[t!]
        \label{gnn-trust-framework}
	\caption{GNN-aided historical reliability evaluation}
	\label{generate implicit trust hypergraph}
		\KwIn {$G^{\text{dir}}$}
            \KwOut{Historical reliability values}
             generate the initial embedding for each device
             \For{$l = 1, \cdots, L$}{
                  \For{$a_j \in {A}$}{
                      \If{ $a_j$ \textnormal{acts as a trustee}}{
                           collect information from its in-degree neighbors using (\ref{w_a_j_get_a_i}), (\ref{ep_a_j_get_a_i}), and (\ref{mu_a_j_get_a_i}) \\
                           aggregate information via the attention mechanism using (\ref{h_a_j_te})
                      }
                      \If{$a_j$ \textnormal{acts as a trustor}}{
                          collect information from its out-degree neighbors using (\ref{omega_a_j_to_a_i}), (\ref{ep_a_j_to_a_i}), and (\ref{mu_a_j_to_a_i}) \\
                          aggregate information via the attention mechanism using (\ref{h_a_j_tr})
                      }
                       combine $a_j$'s embeddings in both roles using (\ref{h_a_j})
                  }
                  \For{$b_m \in {B}$}{
                       collect information from its in-degree neighbors using (\ref{nc_devices_35}), (\ref{ep_devices_36}), and (\ref{mu_b_m_get_a_i}) \\
                       aggregate the propagated information using (\ref{nc_devices_37})
                  }
             }
                  calculate the historical reliability of each terminal device and each EC device using (\ref{h_a_i_rightarrow_a_j}) and (\ref{t_ove_a_i_a_j})
    \end{algorithm}
    
\textbf{Trust fusion of terminal devices in trustor and trustee roles}:
To obtain a more comprehensive embedding for device $a_j$, the embeddings from its roles as both trustor and trustee are merged. This fusion captures both outgoing and incoming trust relationships, reflecting the full trust dynamics of device $a_j$. The fusion is achieved via a fully-connected layer:
\begin{align}
   \label{h_a_j}
    h_{a_j} = \sigma \left( W_{h^{\text{te}}h^{\text{tr}}} \cdot  \left(h_{a_j}^{\text{te}} \parallel h_{a_j}^{\text{tr}} \right)  + b_{h^{\text{te}}h^{\text{tr}}} \right),
\end{align}
where $ W_{h^{\text{te}}h^{\text{tr}}}$ and  $b_{h^{\text{te}}h^{\text{tr}}}$ are the learnable parameters. $h_{a_j}$ is the final embedding of $a_j$ after propagating and aggregating trust information from its neighbors. 

\textbf{Trust propagation and aggregation when EC devices act as trustees}: Embeddings are computed for EC devices, which serve solely as trustees to process tasks from terminal devices. The computation follows the same procedure as for terminal devices acting as trustees, which is given by:
\begin{align}
    \label{nc_devices_35}
     \omega_{b_m \gets a_i} &= W_{b_m \gets a_i} \cdot \chi_{b_m \gets a_i}, \\
      \label{ep_devices_36}
     \epsilon_{b_m \gets a_i} &= W_{b_m \Leftarrow a_i} \cdot \eta_{b_m \gets a_i}, \\
     \label{mu_b_m_get_a_i}
    \mu_{b_m \gets a_i}  &= h_{a_i} \parallel \omega_{b_m \gets a_i} \parallel \epsilon_{b_m \gets a_i}, \\
    \label{nc_devices_37}
    h_{b_m} &= \text{Aggregation} \left(\mu_{b_m \gets a_i}, a_i \in {\mathcal{N}}^{\text{in}}_{b_m}  \right),
\end{align}
where $W_{b_m \gets a_i}, W_{b_m \Leftarrow a_i}$ are the learnable matrices, $\chi_{b_m \gets a_i}$ is the embedding of direct trust value $T^{\text{dir}}_{(a_i,b_m)}$ that device $a_i$ places in device $b_m$, $\eta_{b_m \gets a_i}$ is the embedding of the normalized collaboration frequency $N_{(a_i,b_m)}$, and $\mathcal{N}^{\text{in}}_{b_m}$ is the set of in-degree neighbours of device $b_m$. $h_{b_m}$ is the final embedding of $b_m$, obtained after propagating and aggregating trust information from its neighbours. 

{\textit{2) High-Order Trust Propagation and Aggregation}

After obtaining the device embeddings from first-order neighbours, $L$ propagation and aggregation layers are stacked to enable devices to integrate trust information from their $L$-hop neighbours. In the $l$-th layer, the embedding of device $a_j$ as a trustee is updated as follows:
\begin{align}
    \omega_{a_j \gets a_i}^{(l)} &= W_{a_j \gets a_i}^{(l)} \cdot \chi_{a_j \gets a_i}, \\
    \epsilon_{a_j \gets a_i}^{(l)} &= W_{a_j \Leftarrow a_i}^{(l)} \cdot \eta_{a_j \gets a_i},\\ 
    \mu_{a_j \gets a_i}^{(l)}  &= h_{a_i}^{(l-1)} \parallel \omega_{a_j \gets a_i}^{(l)} \parallel \epsilon_{a_j \gets a_i}^{(l)}, \\
     h_{a_j}^{\text{te}(l)} &= \text{Aggregation} \left(\mu_{a_j \gets a_i}^{(l)}, a_i \in {\mathcal{N}}^{\text{in}}_{a_j} \right). 
\end{align}
Similarly, the embedding $h_{a_j}^{\text{tr}(l)}$ of $a_j$ as a trustor in the $l$-th layer can be obtained through the same operations. Through a fully-connected layer, the comprehensive embedding of device $a_j$ in the $l$-th layer can be calculated as:
\begin{align}
    h_{a_j}^{(l)} = \sigma \left( W^{(l)}_{h^{\text{te}(l)}h^{\text{tr}}} \cdot  \left(h_{a_j}^{\text{te}(l)} \parallel h_{a_j}^{\text{tr}(l)} \right)  + b^{(l)}_{h^{\text{te}}h^{\text{tr}}} \right).
\end{align}
The embedding of $b_m$ in the $l$-th layer, denoted as $h^{(l)}_{b_m}$, is obtained using the same operation as in (\ref{nc_devices_35})--(\ref{nc_devices_37}). Finally, the propagation and aggregation layer outputs the final embeddings for each terminal device and each EC device, denoted as $h_{a_j}^{(L)}$ and $ h_{b_m}^{(L)}$, respectively.
These embeddings incorporate local topological information while also aggregating trust information from $L$-hop neighbours. Specifically, each terminal device's embedding $h_{a_j}^{(L)}$ captures the combined effects of its roles as both trustor and trustee.

\vspace{-0.15 in}
\subsection{Historical Reliability Calculation}  
 To predict trust values between any two devices, we employ the multi-layer perceptron (MLP) as the prediction model. The embeddings of the paired devices are concatenated and fed into the MLP. The historical reliability of device $a_j$ evaluated by device $a_i$ is calculated as: 
 \begin{align}
    \label{h_a_i_rightarrow_a_j}
    h_{(a_i, a_j)}  &= \sigma \left(\text{MLP} \left( h^{(L)}_{a_i} \parallel h^{(L)}_{a_j}  \right) \right),\\
    \label{t_ove_a_i_a_j}
    T^{\text{his}}_{(a_i,a_j)} &= h_{(a_i, a_j)} \left( \argmax \left( h_{(a_i, a_j)}\right) \right), 
 \end{align}
where $\argmax \left( h_{(a_i, a_j)}\right)$ returns the index of the highest value in $h_{(a_i, a_j)}$. Similarly, the historical reliability $T^{\text{his}}_{(a_i,b_m)}$ of device $b_m$ can also be obtained. To train the GNN model, a cross-entropy loss function is used to measure the difference between the predicted values and the ground-truth values. The objective function is formulated as:
    \begin{align}
    \hspace{-0.105 in} \mathcal{L} = &- \frac{1}{|{W}^{\text{dir}}|} \sum_{{ T^{\text{dir}}_{(a_i,a_j)}} \in W^{\text{dir}}} \log \left(h_{(a_i, a_j), T^{\text{dir}}_{(a_i,a_j)}}\right) \nonumber \\ &+ \lambda \parallel \Theta \parallel_2^2,
\end{align}
where $\Theta$ denotes all trainable model parameters, and $\lambda$ controls the $L_2$ regularization strength to prevent over-fitting.

\subsection{Removing Devices Below Trust Thresholds}
Based on their historical reliability, devices that do not meet the trust thresholds are removed from the current network topology $G^{\text{top}}$. 
As indicated in Eq.~(\ref{final_a_i_a_j}), if a terminal device $a_j$'s $T^{\text{his}}_{(a_i,a_j)}$ is lower than $c^{\text{TF}}$, its $T_{(a_i,a_j)}$ will also fall below $c^{\text{TF}}$, making further resource evaluation unnecessary. The same procedure is applied to EC devices based on Eq.~(\ref{final_a_i_b_m}). It is important to note that only devices meeting the trust thresholds are considered for resource evaluation. After removing all devices that do not meet the trust thresholds, a new network topology $G^{\text{new}} = ((A^{\text{new}}, B^{\text{new}}), E^{\text{new}})$ is formed, where $A^{\text{new}}$ is the set of terminal devices with $T^{\text{his}}_{(a_i,a_j)} \geq c^{\text{TF}}$, $ B^{\text{new}}$ is the set of EC devices with $T^{\text{his}}_{(a_i,b_m)} \geq c^{\text{EC}}$, and $E^{\text{new}}$ is the set of communication links connecting these devices.

\section{Agentic AI for Resource Trust Evaluation and Multi-Hop Path Planning}
\label{section_GAI}

Devices identified in the historical reliability evaluation phase are further assessed for their resource trustworthiness with respect to the upcoming task $C$. Due to the dynamic nature of resources and privacy concerns, devices should avoid exposing their resource information to external entities. Consequently, it is essential to equip devices with autonomous decision-making capabilities, enabling them to independently perform task-specific trust evaluation for their resources. The advanced autonomy, environmental adaptability, in-context learning, and multi-agent collaboration capabilities make LAM-enabled agentic AI a promising tool for addressing this challenge.

In this section, we introduce the LAM-enabled agentic AI system designed within the GADAI framework to support resource trust evaluation and cooperative path planning. Each device is equipped with an LAM-powered agent capable of autonomously assessing the trustworthiness of its resources. Building upon these evaluations, the multi-hop path planning problem is decomposed into smaller sub-tasks, which are distributed across multiple agents. Through collaborative decision-making, agents collectively determine the value-maximizing multi-hop path from the task owner to a potential EC device. We first provide a detailed discussion on the construction of the agentic AI system, followed by an examination of resource trust evaluation and distributed multi-hop path planning.

\vspace{-0.1 in}
\subsection{Construction of Agentic AI System}
The agentic AI system comprises three types of LAM-enabled agents, which are implemented using OpenAI-o3-mini and DeepSeek-R1.
 
\textbf{Master agent}: It is responsible for converting the network topology $G^{\text{new}}$ into a textual format and providing device agents with topology information, task details, algorithms, and other relevant data to guide their actions and decisions.

\textbf{Device agent}: Each device is equipped with a device agent responsible for autonomous resource trust evaluation. Each device agent interacts with others through the device communication module. For example, device $a_j$'s agent perceives its own resource status and autonomously determines resource trustworthiness for incoming tasks. The LAM-enabled resource evaluation is detailed in Section~\ref{resource_evaluation}. If device $a_j$ is deemed trustworthy, its agent computes the average VoC according to the received algorithms. Only VoC information is exchanged among neighbouring agents, ensuring that no resource data is shared and privacy is preserved.

\textbf{Algorithm agent}: It determines suitable algorithms for the given tasks and sends them to the master agent, ensuring decisions follow the predefined optimization strategy. The algorithm agent can store predefined algorithms or generate new ones via self-learning or external information integration. In this work, predefined algorithms for value calculation in terminal and edge computing devices are employed.

\subsection{Task-Specific Resource Trust Evaluation and Task-Oriented Multi-Hop Path Planning}
\label{resource_evaluation}

Each device agent collects data on its resource status and intent to establish its initial state. Based on the textual representation of $G^{\text{new}}$, the master agent extracts neighbouring device information. The task owner $a_i$ sends task $C$ and the multi-hop path planning problem to the master agent, which formulates the problem as a query. This query is sent to the algorithm agent and matched to the appropriate value computation algorithm $calculate\_value()$, which is composed of the expressions in Eqs.(\ref{t_rec})--(\ref{average_value}). The selected algorithm is then returned to the master agent. Subsequently, the master agent distributes task $C$, neighbour device data, and the algorithm to each device agent.  Device agents then exchange VoC information with their neighbours to collaboratively plan a multi-hop path that maximizes the average VoC. Using device $a_j$ as an example, its agent consists of the following six core modules: 

\textbf{State}: It includes device $a_j$'s current computational resources, disk space, network resources, device health status, available storage, cooperation willingness, etc.

\textbf{Resource trust evaluation}: 
The advanced reasoning models OpenAI-o3‑mini and DeepSeek-R1 are employed to perform resource evaluation, and a comparison of their performance is presented in the experimental section. To improve the accuracy of evaluation, techniques such as few-shot learning and self-consistency are adopted to enhance the models’ reasoning capabilities~\cite{chain_of_trsut}. Prompts are designed according to the specific characteristics of OpenAI-o3‑mini and DeepSeek-R1 to generate more accurate outputs~\cite{itu6g2023}.
For instance, based on OpenAI’s official documentation, prompts for the o-series models should adhere to the following guidelines~\cite{openai}: \textit{1) keep prompts simple and direct}: write straightforward and concise prompts, as direct instructions yield the best results with o-series models; \textit{2) avoid chain-of-thought prompts}: since these models perform reasoning internally, instructing them to ``think step by step" or ``explain your reasoning" is unnecessary; \textit{3) use delimiters for clarity}: break complex prompts into sections using delimiters like Markdown, XML tags, or quotes. This structured format enhances model accuracy; \textit{4) show rather than tell}: for complex outputs, providing a few examples of inputs and expected outputs can be beneficial. Based on these guidelines, prompts are constructed to evaluate the trustworthiness of each device, as illustrated in Fig.~\ref{task_resource_analysis}. For each terminal device $a_j$, the resource trust $T^{\text{res}}_{a_i,a_j}$ is set to 1 if its resources meet the requirements of task $C$, and 0 otherwise. The same procedure is applied to EC devices.

\textbf{Memory}: It stores device $a_j$'s neighbours, the maximum average VoC $\overline{V}^{\text{max}}_{(a_i,a_j)}$ and hop count $k$ from the task owner $a_i$ to itself, as well as the preceding device on the path.

\begin{figure}[t!]
\centering
\includegraphics[scale=0.95]{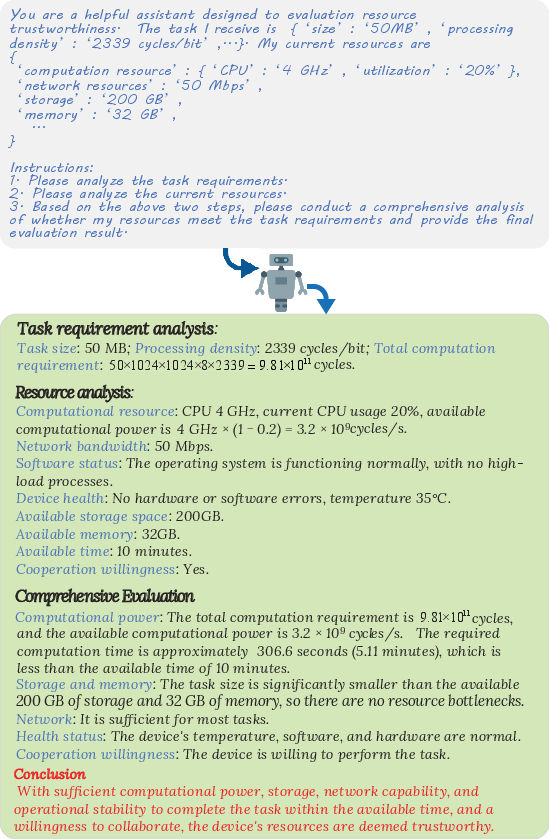}
\caption{Task-specific resource trust evaluation for an EC device using OpenAI-o3-mini.}
\label{task_resource_analysis}
\end{figure}

\textbf{Send}: If device $a_j$ is trusted, its agent broadcasts the currently known $\overline{V}^{\text{max}}_{(a_i,a_j)}$ to its neighbours; otherwise, its agent sends no messages. 

\textbf{Update}: Upon receiving $\overline{V}^{\text{max}}_{(a_i,a_r)}$ from its neighbor $a_r$, device $a_j$ calculates a new average VoC $\overline{V}_{(a_i,a_j)}$, based on the received value. If the calculated $\overline{V}_{(a_i,a_j)}$ is greater than the currently stored maximum average VoC $\overline{V}^{\text{max}}_{(a_i,a_j)}$, device $a_j$ updates $\overline{V}^{\text{max}}_{(a_i,a_j)}$ with the new $\overline{V}_{(a_i,a_j)}$. Subsequently, it broadcasts the updated $\overline{V}^{\text{max}}_{(a_i,a_j)}$ to all its neighbors.

\textbf{Termination}: Iteration of the send and update phases ceases when the  
$\overline{V}^{\text{max}}_{(a_i,a_j)}$ values for all devices converge.

Each EC device $b_m$ receives multiple path messages and selects the path $\pi_{(a_i,b_m)}$ with the maximum average VoC. It then constructs this path by tracing back to the task owner $a_i$, utilizing the preceding device recorded at each hop. After the task owner $a_i$ receives the optimal path messages from all available EC devices, it chooses the path with the highest average VoC value as the final collaboration path. Task $C$ is subsequently forwarded along the devices on this path, reaching the last-hop EC device for execution.

\subsection{Complexity Analysis}
\textit{1) Computational complexity analysis}: 
The computational complexity of the proposed GADAI framework can be divided into three components: historical collaboration graph construction, GNN-assisted historical reliability evaluation, and agentic AI-enabled resource evaluation and multi-hop path planning. For historical collaboration graph construction, generating $G^{\text{dir}}$ requires processing both devices and historical collaboration data, resulting in a computational complexity of $O(|A| + |B| + |D|)$, where $|A|$, $|B|$, and $|D|$ denote the numbers of terminal devices, EC devices, and recorded historical collaborations, respectively. In GNN-assisted reliability evaluation, trust information is propagated and aggregated across the graph $G^{\text{dir}}$. Considering $L$ GNN layers, the total complexity is $O(LQ(|E^{\text{dir}}|H_a^2 + (|A|+|B|)H_a^2 + |E^{\text{dir}}|H_aH_T))$~\cite{Wu2021Comprehensive}, where $Q$ is the number of attention heads, $|E^{\text{dir}}|$ is the number of edges in $G^{\text{dir}}$, $H_a$ is the dimension of device embeddings, and $H_T$ is the dimension of trust embeddings. For multi-hop path planning, the time complexity is at most $O((|A^{\text{new}}| + |B^{\text{new}}|)|E^{\text{new}}|)$, where $|A^{\text{new}}|$, $|B^{\text{new}}|$, and $|E^{\text{new}}|$ are the numbers of terminal devices, EC devices, and edges in the network topology $G^{\text{new}}$, respectively.

\textit{2) Space complexity analysis}: The space complexity of the proposed framework can be analyzed for each component. The space complexity for generating the graph $G^{\text{dir}}$ is $O(|A| + |B| + |E^{\text{dir}}|)$. The space complexity in the GNN component is $O(LQ(H_a^2 + H_aH_T + |E^{\text{dir}}|) + (|A| + |B|)H_a + |E^{\text{dir}}|H_T)$~\cite{Wu2021Comprehensive}. For multi-hop path planning, the space complexity is $O(|A^{\text{new}}| + |B^{\text{new}}| + |E^{\text{new}}|)$.

\section{Experimental Results}
\label{section_experiment}

\subsection{Experimental Setup}
Two task types are considered—face recognition and virus scanning—with their default features summarized in Table~\ref{task_feature}. The iPhone 15 and Pixel 8 are used for task forwarding, and their performance data is collected to construct two terminal device models. A Lambda Vector workstation is employed for task computation, and its performance is measured to construct an EC device model. The default parameters for these devices are listed in Table~\ref{device_feature}. It is important to note that these values may fluctuate during runtime. We use the discrete-event network simulator NS-3 and Python bindings to construct a Wi-Fi-based cooperative system, which includes 100 iPhone 15 device models, 100 Pixel 8 device models, and 10 Lambda device models. The service prices of iPhone 15, Pixel 8, and Lambda are 0.02~$\$/$second, 0.01~$\$/$second, and 0.002~$\$/$second, respectively. The channel bandwidth is set to 5 MHz. The transmit power of all terminal devices is set as $100$ mW, and the noise power is -80 dBm. $\alpha_1$ and $\alpha_2$ are set to 0.6 and 0.4, respectively. A total of 5,000 executions are performed for the two types of tasks. The initialized embeddings of devices are set to a dimension of 128. Following~\cite{9772384}, the GNN model consists of $L = 3$ propagation and aggregation layers, with output dimensions of 32, 64, and 32 for the first, second, and third layers, respectively. In terms of hyperparameters, the learning rate is chosen from $\{10^{-1}, 10^{-2}, 10^{-3}, 10^{-4}\}$, the $L_2$ regularization coefficient from $\{10^{-5}, 10^{-4}\}$, and the dropout rate from $\{0, 0.1, 0.3, 0.5, 0.8 \}$. The Xavier initializer is used to initialize the GNN parameters. The dataset is split into 80\% training and 20\% testing subsets. Within the training set, 5-fold cross-validation is employed, and early stopping is used to prevent overfitting. The GNN model is trained on the Lambda Vector workstation. Without specification, the reported experiment results correspond to a learning rate of $10^{-2}$, an $L_2$ regularization coefficient of $10^{-5}$, and a dropout rate of $0$, as these settings yield the best performance.

\begin{table}[!t] 
    \footnotesize  
    \centering
    \renewcommand{\arraystretch}{1.6} 
    \caption{Task Features}
    \label{task_feature}
    \begin{tabular}{m{1.3cm}<{\centering}|m{0.4cm}<{\centering}|m{2.3cm}<{\centering}|m{0.7cm}<{\centering}|m{0.7cm}<{\centering}|m{0.9cm}<{\centering}}
        \hline 
        Task & Size & Processing density & $s^{\text{TF}}_{\text{soft}}$, $s^{\text{TF}}_{\text{hard}}$ & $s^{\text{EC}}_{\text{soft}}$, $s^{\text{EC}}_{\text{hard}}$ & $c^{\text{TF}}$, $c^{\text{EC}}$ \\
        \hline
        \hline
        Face recognition & 200 MB & 2,339 cycles/bit~\cite{7264984} & \$1,~\$2 & \$2,~\$5 & 0.2, 0.2 \\
        \hline
        Virus Scanning & 200 MB & 32,946 cycles/bit~\cite{7264984}, \cite{chun2011clonecloud} & \$1,~\$2 & \$2,~\$5 & 0.2, 0.2 \\
        \hline
    \end{tabular}
\end{table}

\begin{table}[!t] 
    \footnotesize  
    \centering
    \renewcommand{\arraystretch}{1.3}
    \caption{Device Parameters}
    \label{device_feature}
    \begin{tabular}{p{1.75cm}<{\centering}|p{0.95cm}<{\centering}|p{0.95cm}<{\centering}|p{1.15cm}<{\centering}|p{1.9cm}<{\centering}}
        \hline 
        Device & Memory & Storage & CPU & GPU \\
        \hline
        \hline
        iPhone 15 & 6 GB & 128 GB & 3.46 GHz &  / \\
        \hline
        Pixel 8 &  8 GB & 128 GB  & 2.91 GHz  & / \\
        \hline
        Lambda Vector & 32 GB & 3.84 TB & 4 GHz & NVIDIA A6000\\
          \hline 
    \end{tabular}
\end{table}

The experimental data processing involves two main stages. In the first stage, device models in NS-3 are constructed from real measurements. This involves extracting statistical features from devices, mapping them to NS-3 node parameters, and simulating task forwarding and computation to collect inter-node collaboration data. In the second stage, the collected NS-3 data are cleaned to address missing values and anomalies. Subsequently, node and edge features are extracted, the graph structure is constructed, and all features are normalized and encoded to form the training and testing datasets.


\begin{figure}[!t]
\centering
\includegraphics[scale=0.9]{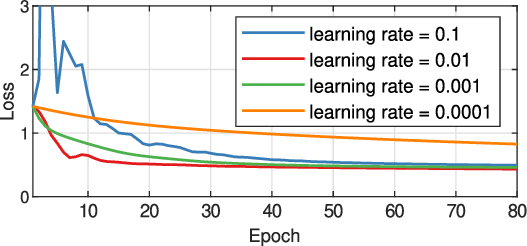}
\caption{Training loss comparison for different learning rates.}
\label{loss}
\end{figure}

\subsection{Comparative Analysis of Historical Reliability Evaluation}

In this section, the GNN-aided historical reliability evaluation in the proposed GADAI framework is investigated. The convergence performance of the GNN model is evaluated first under different learning rates. As shown in Fig.~\ref{loss}, a learning rate of 0.1 leads to unstable convergence with large fluctuations. A learning rate of 0.01 achieves the fastest and most stable convergence. When the learning rate is 0.001, the loss decreases smoothly but more slowly. At a learning rate of 0.0001, the convergence is the slowest and remains at a high loss of 0.8230. These results indicate that 0.01 provides the best balance between convergence speed and stability.

NeuralWalk~\cite{8737469}, originally designed for trust assessment in social networks, is used for comparison. Two widely used metrics--Root Mean Square Error (RMSE) and Mean Absolute Error (MAE)--are employed to evaluate the accuracy of trust assessment. Lower values of RMSE and MAE indicate higher accuracy. The proportions of the training set are set to 80\%, 60\%, and 40\% of the entire dataset. All reported results are averaged over 10 runs. As shown in Table~\ref{accuracy_analysis}, when the training set ratio is 80\%, GADAI achieves its best MAE score of 0.081, outperforming NeuralWalk, and its RMSE is also significantly lower than that of NeuralWalk. As the proportion of the training set decreases, GADAI’s MAE and RMSE scores show a slight decline because of the reduced amount of training data. Nevertheless, GADAI consistently maintain superior performance compared to NeuralWalk across all training set sizes. This indicates that GADAI is more effective in capturing the mutual influence among devices, as well as in aggregating and propagating trust information across devices.

\begin{table}[!t] 
    \footnotesize  
    \centering
    \renewcommand{\arraystretch}{1.3}
    \caption{Accuracy Analysis under Different Training Set Sizes}
    \label{accuracy_analysis}
    \begin{tabular}{p{1.8cm}<{\centering}|p{1.6cm}<{\centering}|p{1.8cm}<{\centering}|p{1.8cm}<{\centering}}
        \hline 
        {Method} & {Training set} & RMSE & MAE \\
        \hline
        \hline
        \multirow{3}{*}{GADAI (GNN)} & 80\% & 0.107 $\pm$ 0.003  & 0.081 $\pm$ 0.002\\
        \cline{2-4}
         & 60\% & 0.112  $\pm$ 0.004  & 0.084 $\pm$ 0.002 \\
         \cline{2-4}
          & 40\% &  0.119  $\pm$ 0.004 & 0.090 $\pm$ 0.003 \\
        \hline
         \multirow{3}{*}{NeuralWalk} & 80\% &  0.153 $\pm$ 0.003 & 0.115 $\pm$ 0.003 \\
        \cline{2-4}
         & 60\% &  0.159 $\pm$ 0.004 & 0.120 $\pm$ 0.003 \\
         \cline{2-4}  
         & 40\% &  0.167 $\pm$ 0.004 & 0.126 $\pm$ 0.003 \\
          \hline
    \end{tabular}
\end{table}

\begin{table}[!t] 
    \footnotesize  
    \centering
    \renewcommand{\arraystretch}{1.3}
    \caption{Accuracy Analysis under Different Layer Dimensions}
    \label{layer_comparison}
    \begin{tabular}{m{3cm}<{\centering}|m{2cm}<{\centering}|m{2cm}<{\centering}}
        \hline 
         Layer dimensions & RMSE & MAE \\
        \hline
        \hline
        16, 32, 16 & 0.125 & 0.092 \\
        \hline
        32, 64, 32 & 0.107 & 0.081  \\
        \hline
        64, 128, 64 & 0.115 & 0.087 \\
        \hline
    \end{tabular}
\end{table}

\begin{table}[!t] 
    \footnotesize  
    \centering
    \renewcommand{\arraystretch}{1.3}
    \caption{Accuracy analysis under different head numbers}
    \label{head_comparison}
    \begin{tabular}{p{1.5cm}<{\centering}|p{0.9cm}<{\centering}|p{0.9cm}<{\centering}|p{0.9cm}<{\centering}|p{0.9cm}<{\centering}|p{0.9cm}<{\centering}}
        \hline 
         & $Q$ = 1 & $Q$ = 2 & $Q$ = 3 & $Q$ = 4 & $Q$ = 5 \\
         \hline
        \hline
         MAE & 0.084  & 0.081 & 0.083 & 0.085 & 0.089 \\
        \hline
    \end{tabular}
\end{table}

\begin{table}[!t] 
    \footnotesize  
    \centering
    \renewcommand{\arraystretch}{1.3}
    \caption{Comparison of Resource Evaluation of Terminal Devices Using Different LAMs}
    \label{tf_resource_trust}
    \begin{tabular}{m{1.3cm}<{\centering}|m{2.2cm}<{\centering}|m{1.5cm}<{\centering}|m{1.8cm}<{\centering}}
        \hline 
        {Terminal device} & {Model} & Accuracy ($T^{\text{res}}_{a_i,a_j}$) & Inference time (second) \\
        \hline
        \hline
        \multirow{2}{*}{iPhone 15} & OpenAI-o3-mini & 100\% & 1.35 \\
        \cline{2-4}
         & DeepSeek-R1:1.5B & 99\% & 2.62 \\
        \hline
         \multirow{2}{*}{Pixel 8} & OpenAI-o3-mini &  100\% & 1.37 \\
        \cline{2-4}
         & DeepSeek-R1:1.5B & 98.6\% & 3.01 \\
          \hline
    \end{tabular}
\end{table}

\begin{table}[!h] 
    \footnotesize  
    \centering
    \renewcommand{\arraystretch}{1.3}
    \caption{Comparison of Resource Evaluation of EC Device Using Different LAMs}
    \label{ec_resource_trust}
    \begin{tabular}{m{1.3cm}<{\centering}|m{2.2cm}<{\centering}|m{1.3cm}<{\centering}|m{1.8cm}<{\centering}}
        \hline 
       {EC Device} & {Model} & Accuracy ($T^{\text{res}}_{a_i,b_m}$) & Inference time (second) \\
          \hline
          \hline
          \multirow{3}{*}{Lambda} & OpenAI-o3-mini & 99.3\% & 3.22\\
        \cline{2-4}
         & DeepSeek-R1:1.5B &  69.3\% &  2.68\\
          \cline{2-4}
         & DeepSeek-R1:14B & 90.6\% &  8.39\\
         \hline
    \end{tabular}
\end{table}

 We further investigate the impact of hidden layer dimensions on model performance. As shown in Table \ref{layer_comparison}, increasing the dimensions from (16, 32, 16) to (32, 64, 32) produces lower values of MAE and RMSE. However, further increasing the dimensions to (64, 128, 64) provides no additional performance gains. Considering the trade-off between performance and computational cost, we chose the (32, 64, 32) configuration in this paper.

We also examine the effect of the number of attention heads. As shown in Table~\ref{head_comparison}, MAE reaches its minimum at $Q = 2$, indicating that two attention heads provide the optimal result. Increasing the number of attention heads may lead to overfitting, thereby degrading performance.

\subsection{Comparative Analysis of Resource Trust Evaluation}
The performance of the proposed GADAI on resource trust evaluation is examined in this subsection. Unless otherwise stated, the results presented in the following subsections are based on face recognition tasks. Agents are implemented using both OpenAI-o3-mini~\cite{openai} and DeepSeek R1~\cite{itu6g2023} models for performance comparison. Since OpenAI-o3-mini is a closed-source model that cannot be deployed locally, it is accessed exclusively via OpenAI's APIs, meaning all computations are performed remotely on OpenAI's servers. For comparison, DeepSeek R1 models are locally deployed on devices to evaluate on-device inference performance. Considering differences in device capabilities, DeepSeek-R1:1.5B (1.1 GB) is deployed on terminal devices, while both DeepSeek-R1:1.5B and DeepSeek-R1:14B (9.0 GB) are deployed on EC devices. All results are reported over 300 runs.

Table~\ref{tf_resource_trust} presents the resource trust evaluation performance of different LAM models on terminal devices. Since terminal devices handle task forwarding, their resource trust evaluation is relatively straightforward, involving simple logical checks such as their available storage and willingness to forward tasks. OpenAI-o3-mini achieves 100\% accuracy on both types of terminal devices, which can be attributed to its strong reasoning capability that is sufficient for resource trust evaluation. Furthermore, OpenAI-o3-mini takes lower inference time because computations are executed on OpenAI’s high-performance servers rather than on terminal devices. Despite its smaller model size, DeepSeek-R1:1.5B exhibits high accuracy. This indicates that it is well-suited for the basic logical verification tasks required in resource trust evaluation on terminal devices. Its inference time is, however, longer than that of OpenAI-o3-mini, mainly because of the limited computational resources available on terminal devices.

Table~\ref{ec_resource_trust} shows the resource trust evaluation on EC devices. Evaluating resource trust for EC devices is more complex than for terminal devices, as it involves computing operations related to communication and computational resources, as shown in Fig.~\ref{task_resource_analysis}. OpenAI-o3-mini achieves high accuracy due to its strong reasoning ability and extensive knowledge, enabling it to handle the complex evaluation of EC resources effectively. In contrast, DeepSeek-R1:1.5B exhibits lower accuracy, as it struggles to address the complexities of resource trust evaluation for EC devices. Compared with DeepSeek-R1:1.5B, DeepSeek-R1:14B achieves higher accuracy due to its larger model size and stronger reasoning capability. This improvement, however, comes at the cost of increased inference time.

\begin{figure}[!t]
      \centering
      \subfigure[]{\includegraphics[scale=0.455]{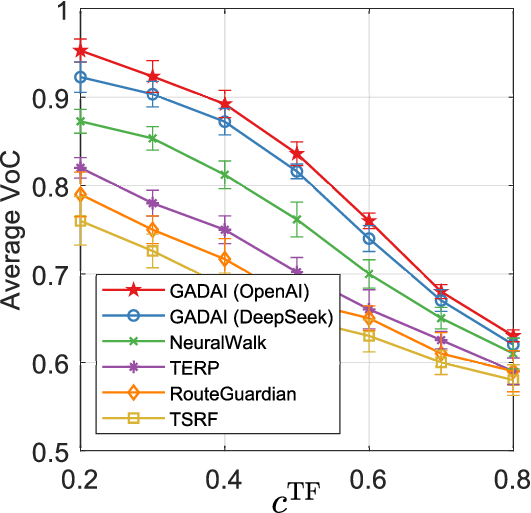}}
      \hspace{-0.01 in}\subfigure[]{\includegraphics[scale=0.455]{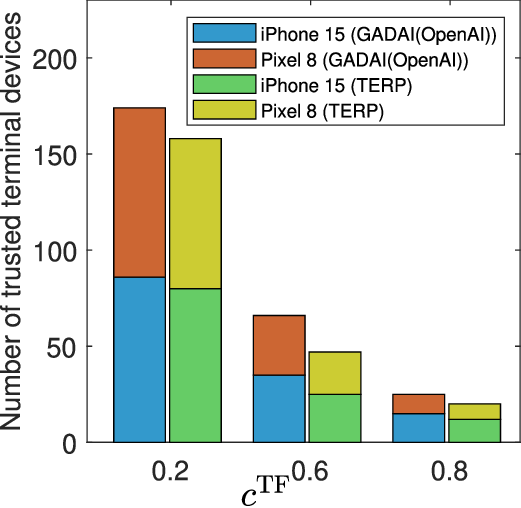}}
      \caption{Comparison of results with varying minimum trust demand $c^{\text{TF}}$. (a)~Comparison of the average VoC values. As $c^{\text{TF}}$ increases, the average VoC values decrease for all methods, while GADAI consistently achieves higher values than the comparison methods. (b)~Comparison of the number of trusted terminal devices. As $c^{\text{TF}}$ increases, the number of trusted terminal devices for all methods decreases.}
     \label{c_tf}
\end{figure}

\begin{figure}[!t]
      \centering
      \subfigure[]{\includegraphics[scale=0.455]{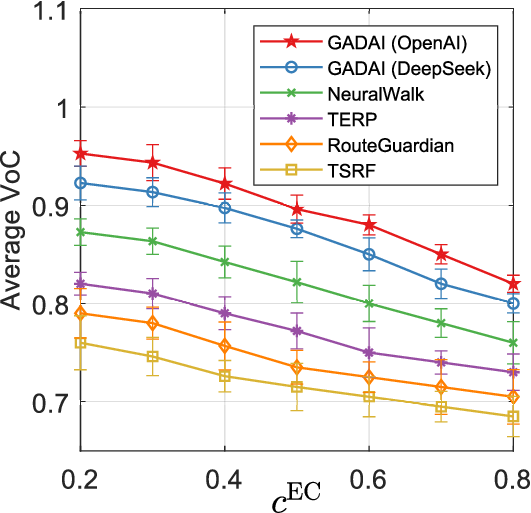}}
      \hspace{-0.03 in}
      \subfigure[]{\includegraphics[scale=0.455]{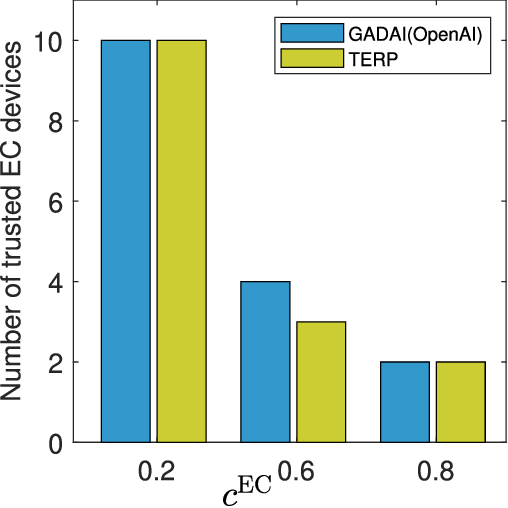}}
     \caption{Comparison of results under varying the minimum trust demand $c^{\text{EC}}$. (a)~Comparison of the average VoC values. As $c^{\text{EC}}$ rises, all methods experience a decline in the average VoC values; however, GADAI consistently maintains superior values. (b)~Comparison of the number of trusted EC devices. As $c^{\text{EC}}$ increases, the number of trusted EC devices for all methods decreases.}
    \label{c_ec}
\end{figure}

\subsection{Impact of Minimum Trust Thresholds on Average VoC}
This subsection investigates the impact of the minimum trust thresholds, i.e., $c^{\text{TF}}$ and $c^{\text{EC}}$, of task $C$ on the average VoC and the number of trusted terminal devices. A comparative analysis is conducted with RouteGuardian~\cite{7986943}, TSRF~\cite{doi:10.1155/2014/209436}, TERP~\cite{7194748}, and NeuralWalk. Since NeuralWalk is applicable only to historical reliability evaluation, the resource evaluation and multi-hop path planning approaches proposed in this paper are employed when utilizing NeuralWalk. The mean results are presented along with standard deviation (SD) bars to provide a quantitative measure of result dispersion. These SD bars indicate the standard deviation over 10 independent runs. In Fig.~\ref{c_tf}~(a), it can be seen that as $c^{\text{TF}}$ increases, the average VoC values obtained by all algorithms decrease. This is because the number of terminal devices satisfying the minimum trust threshold declines, reducing the number of available devices and consequently lowering the average VoC values. The results in Fig.~\ref{c_tf}~(b) further support this observation. Moreover, the proposed GADAI method, whether implemented with OpenAI or DeepSeek, consistently achieves higher average VoC values than the comparison algorithms, and the number of trusted terminal devices it identifies is also significantly greater. This advantage stems from GADAI’s ability to comprehensively evaluate historical reliability and task-specific resource trust for devices, enabling more precise selection of available devices.

A further investigation is conducted to evaluate the impact of the minimum trust threshold $c^{\text{EC}}$ on the average VoC and the number of trusted EC devices. As illustrated in Fig.~\ref{c_ec}~(b), increasing $c^{\text{EC}}$ reduces the number of trusted EC devices identified by both methods. This reduction increases the cost of planning multi-hop paths, which in turn lowers the average VoC values, as observed in Fig.~\ref{c_ec}~(a). The average VoC exhibits a slower decreasing trend compared with the results shown in Fig.~\ref{c_tf}~(a). This is because the minimum trust threshold $c^{\text{EC}}$ impacts only the number of EC devices rather than the number of terminal devices. Nevertheless, GADAI continues to outperform the comparison approaches by achieving higher average VoC values.

\begin{figure}[!t]
      \centering
      \subfigure[]{\includegraphics[scale=0.455]{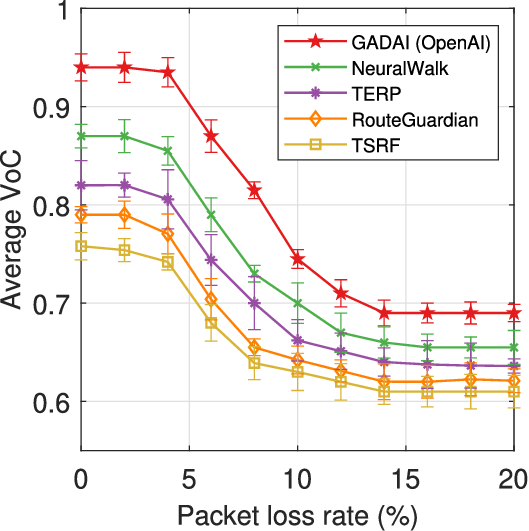}}
      \hspace{-0.03 in}\subfigure[]{\includegraphics[scale=0.455]{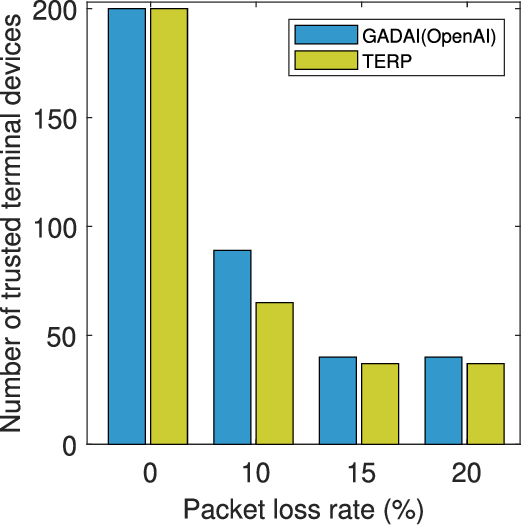}}
      \caption{Comparison of results when varying packet loss rate. (a) Comparison of the average VoC values. Increasing the packet loss rate reduces the average VoC, and GADAI attains higher average VoC values compared with the other methods. (b) Comparison of the number of trusted terminal devices. Increasing the packet loss rate reduces the number of trusted devices.}
     \label{plr}
\end{figure}

\begin{figure}[!t]
      \centering
      \subfigure[]{\includegraphics[scale=0.455]{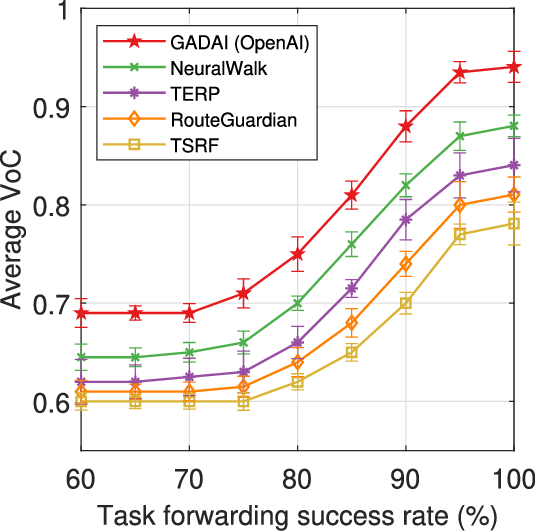}}
      \hspace{-0.03 in}\subfigure[]{\includegraphics[scale=0.455]{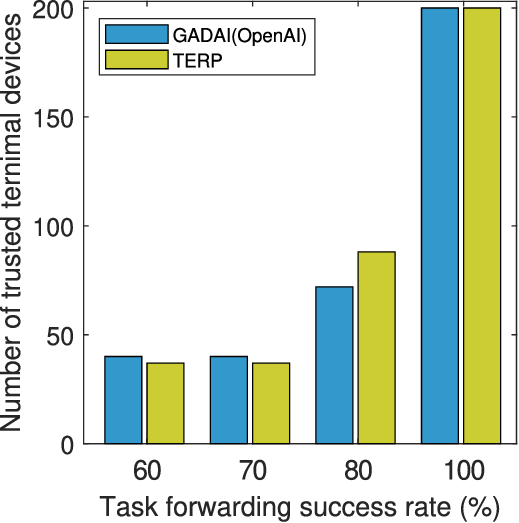}}
      \caption{Comparison of results when varying task forwarding success rate. (a) Comparison of average VoC. As the success rate increases, the average VoC increases, while GADAI consistently achieves higher average VoC values than the other methods. (b) Comparison of the number of trusted terminal devices. The number of trusted terminal devices rises as the success rate increases.}
     \label{tfsr}
\end{figure}

\begin{figure}[t!]
\centering
\includegraphics[scale=0.56]{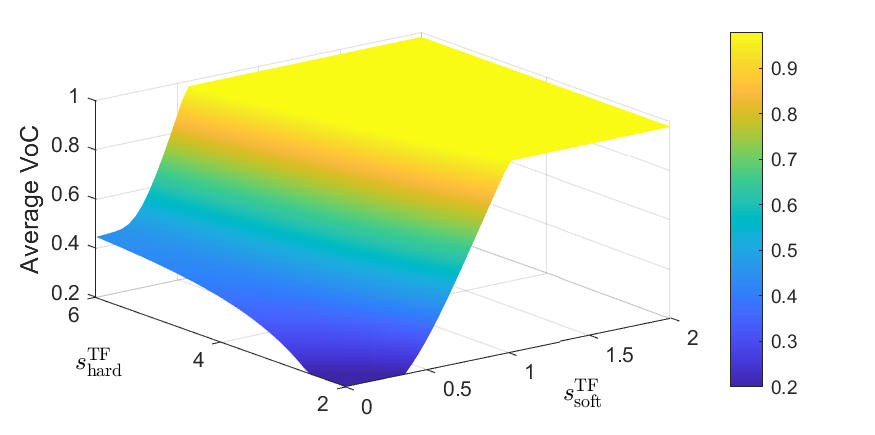}
\caption{The average VoC increases as $s^{\text{TF}}_{\text{soft}}$ and  $s^{\text{TF}}_{\text{hard}}$ increase.}
\label{softhard}
\end{figure}


\subsection{Impact of Packet Loss Rate and Task Forwarding Success Rate on Average VoC}
As defined in Eq. (\ref{a_i_dir_a_j}), the packet loss rate and task forwarding success rate jointly determine the historical reliability of terminal devices. The impact of the packet loss rate is first examined in this subsection. Specifically, 160 out of 200 terminal devices are configured with an identical packet loss rate, and this parameter is varied to assess its influence on performance. As shown in Fig.~\ref{plr}~(a), when the packet loss rate is below 4\%, all approaches achieve high average VoC values. This is because all terminal devices are regarded as trustworthy and thus available to provide task forwarding service. With increasing packet loss rates, untrusted devices are progressively identified, reducing the number of trusted devices and causing the average VoC to drop sharply. The results in Fig.~\ref{plr}~(b) further show that higher packet loss rates lead to fewer trusted devices. When the packet loss rate exceeds 14\%, the average VoC values of all approaches become stable, as all untrusted devices are fully identified. Obviously, the proposed GADAI approach consistently outperforms the comparison algorithms in achieving higher average VoC values.

The impact of the task forwarding success rate on results is investigated. As shown in Fig.~\ref{tfsr}~(a), when the task forwarding success rate falls below 70\%, all approaches yield relatively low and stable average VoC values. This stability occurs because devices with low task forwarding success rates are recognized as untrustworthy and consequently excluded from cooperation, thus exerting no impact on the average VoC. This finding is further supported by Fig.~\ref{tfsr}~(b), where the proposed GADAI model identifies 40 trusted devices at success rates of 60\% and 70\%. As the task forwarding success rate increases, the average VoC values improve as some devices transition from untrustworthy to trustworthy, thus enlarging the pool of available devices. Overall, the experimental results demonstrate that the proposed GADAI approach consistently outperforms the comparison algorithms by achieving higher average VoC values across different success rates.


\subsection{Impact of $s^{\text{TF}}_{\text{soft}}$ and  $s^{\text{TF}}_{\text{hard}}$}
According to Eq. (\ref{votf}), the thresholds $s^{\text{TF}}_{\text{soft}}$ and  $s^{\text{TF}}_{\text{hard}}$ have a significant impact on the VoC of terminal devices. As shown in Fig.~\ref{softhard}, increasing $s^{\text{TF}}_{\text{soft}}$ raises the average VoC by delaying the onset of VoC decay for terminal devices involved in task forwarding. Similarly, increasing $s^{\text{TF}}_{\text{hard}}$ raises the average VoC by postponing the end of VoC decay, thereby reducing occurrences where the VoC of these devices drops to zero.

\begin{figure*}[!t]
      \centering
      \hspace{0 in}\subfigure[]{\includegraphics[scale=0.55]{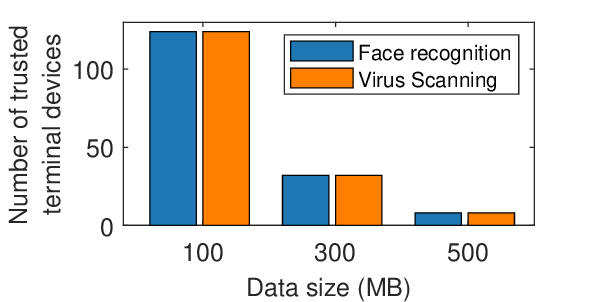}}
      \subfigure[]{\includegraphics[scale=0.55]{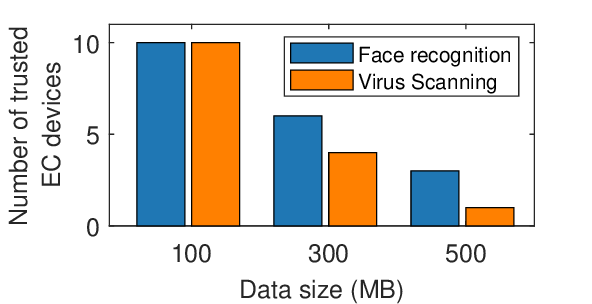}}
      \subfigure[]{\includegraphics[scale=0.55]{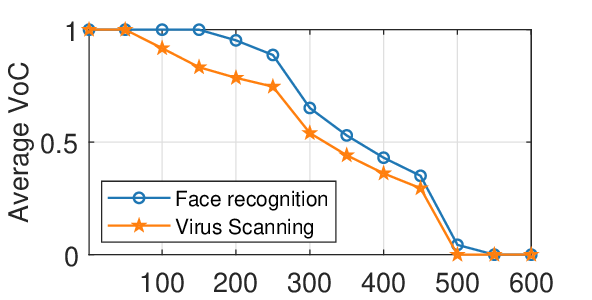}}
      \caption{Comparison of face recognition and virus scanning tasks. (a) Task size affects the resource trustworthiness of terminal devices, and as it increases, the number of trusted terminal devices declines. (b) Task size and processing density influence the resource trustworthiness of EC devices. (c) The average VoC values for both tasks decrease with increasing task size, and the virus scanning task yields lower average VoC values owing to its higher processing density.}
     \label{datasize}
\end{figure*}

\begin{figure}[!t]
      \centering
      \subfigure[]{\includegraphics[scale=0.44]{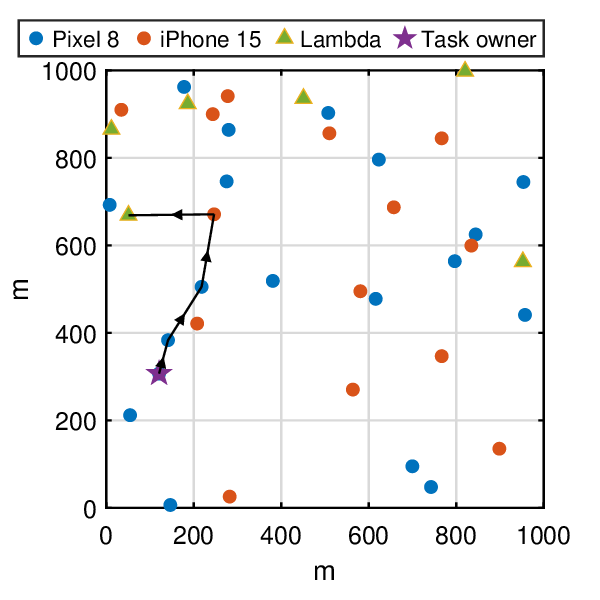}}
      \hspace{-0.1 in}\subfigure[]{\includegraphics[scale=0.44]{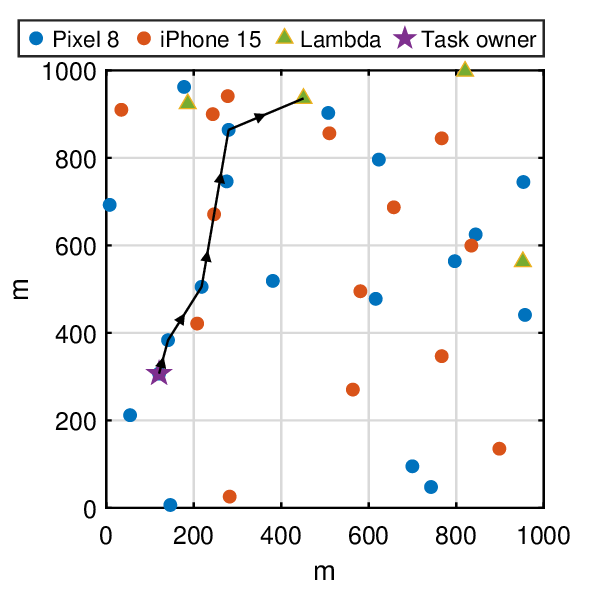}}
      \caption{Comparison of planned multi-hop paths. (a) The face recognition task is forwarded from the task owner to the selected EC device over 4 hops. (b) The virus scanning task is forwarded from the task owner to the selected EC device over 5 hops.}
     \label{path}
\end{figure}

\subsection{Impact of Task Size and Processing Density}
In Definitions 1 and 2, the resources of devices constitute an essential component of their trustworthiness for a given task. Consequently, variations in resource attributes directly affect the trust level of devices. In this subsection, two tasks--face recognition and virus scanning--are used to analyze the impact of task size and processing density on device trustworthiness and the average VoC. As illustrated in Fig.~\ref{datasize}~(a), the number of trusted terminal devices decreases as the task size increases. This reduction occurs because some devices lack sufficient storage capacity to accommodate larger tasks, which makes their resource evaluation results untrustworthy. Furthermore, both tasks yield the same number of trustworthy devices across different task sizes. This finding indicates that processing density has no impact on the resource trustworthiness of terminal devices.

In Fig.~\ref{datasize}~(b), the number of trusted EC devices decreases for both tasks as the task size increases. However, the number of trusted devices for the virus scanning task is noticeably lower than that for the face recognition task. This discrepancy is primarily due to differences in processing density between the two tasks. For the same task size, the virus scanning task requires a longer processing time than the face recognition task according to Eq.~(\ref{processing_time}), because its processing density is 32,946 cycles/bit, which is significantly higher than the 2,339 cycles/bit of the face recognition task. Consequently, some EC devices have insufficient available computational time, resulting in their resources being deemed untrustworthy.

In Fig.~\ref{datasize}~(c), the average VoC of the virus scanning task is significantly lower than that of the face recognition task. This is due to its higher processing density, which requires greater expenditure and thus reduces the average VoC. As the task size increases, the average VoC for both tasks gradually decreases until it reaches zero. This occurs when the task expenditures on both terminal and EC devices exceed their corresponding hard thresholds, $s^{\text{TF}}_{\text{hard}}$ and $s^{\text{EC}}_{\text{hard}}$.

The sizes of both tasks are set to 300 MB, and the planned multi-hop paths are examined. As shown in Fig.~\ref{path}~(a), for the face recognition task, there are six trusted EC devices available, and the task is successfully forwarded from the task owner to the selected EC device over four hops. In Fig.~\ref{path}~(b), for the virus scanning task, the number of trusted EC devices decreases to four, and the task is forwarded from the task owner to the selected EC device over five hops. This demonstrates that task attributes, such as processing density, affect the resource trustworthiness of devices and the planned multi-hop paths.

\section{Conclusions and Future Work}
\label{section_conclusion}
This paper has investigated the problem of collaborator trust evaluation and multi-hop cooperative path planning for maximizing the average VoC. We have proposed the GADAI framework, which implements independent evaluation of multiple trust perspectives and an integrated decision-making mechanism, enhancing the accuracy of trust evaluation. For evaluating historical reliability, the GNN-assisted model is employed to propagate and aggregate trust information across multi-hop neighbours, resulting in more precise reliability evaluations. For dynamic and privacy-sensitive device resources, we have designed the agentic AI system, enabling devices to autonomously perform task-specific resource trust evaluation through LAM-enabled agents. Furthermore, agents collaboratively perform task-oriented multi-hop cooperative path planning in a distributed manner. Experimental results have shown that the proposed GADAI framework outperforms comparison algorithms in both accurate trust evaluation and the planning of value-maximizing multi-hop paths. By integrating multiple trust-related perspectives and enabling distributed collaboration, GADAI provides a more accurate and adaptive trust decision-making framework, thereby enhancing robustness, scalability, and privacy in dynamic environments. Our future research will focus on two directions. First, we plan to design a fully agentic AI-based approach to evaluate trust across multiple dimensions, thereby achieving a truly intelligent assessment mechanism. Second, we will explore techniques such as quantization, pruning, and retrieval-augmented generation to optimize the inference speed and accuracy of LAMs deployed locally.


\footnotesize


\end{document}